\documentstyle[amssymb,aps]{revtex}

\large

\begin{document}
\title{Dynamic scaling in vacancy-mediated disordering.}
\author{B. Schmittmann, R.K.P. Zia and Wannapong Triampo}
\address{Department of Physics \\and \\Center 
for Stochastic Processes in
Science and Engineering, \\
Virginia Tech, Blacksburg, V.A. 24061-0435, USA }
\date{November 23, 1999 }
\maketitle

\begin{abstract}
We consider the disordering dynamics of an interacting binary alloy with a
small admixture of vacancies which mediate atom-atom exchanges. Starting
from a perfectly phase-segregated state, the system is rapidly heated to a
temperature in the disordered phase. A suitable disorder parameter, namely,
the number of broken bonds, is monitored as a function of time. Using
Monte Carlo
simulations and a coarse-grained field theory, we show that the late stages
of this process exhibit dynamic scaling, characterized by a set of scaling
functions and exponents. We discuss the universality of these exponents and
comment on some subtleties in the early stages of the disordering process.
\end{abstract}


\vspace{5mm}


\section{Introduction}

The dynamics of binary alloys, undergoing mixing or segregation processes,
is a venerable problem in the physics and materials science communities \cite
{reviews}. Much interest has focused specifically on the dynamics of phase
ordering and domain growth, following a rapid quench below the coexistence
curve. Starting from an initial non-equilibrium configuration, the system
evolves towards its final equilibrium state, giving rise to fundamental
questions in both disciplines. Crucial to metallurgy, the domain morphology
is intimately linked to structural stability while statistical physics
focuses on the universal, self-similar aspects of its temporal evolution,
manifested in characteristic growth exponents and dynamic scaling.

A key ingredient in the study of these processes is the mechanism by which
two particles exchange positions. From the viewpoint of modelling and
simulation purposes, {\em direct} particle-particle exchanges obviously lead
to the simplest codes and coarse-grained equations. Moreover, invoking
universality, it is expected that the details of the microscopic mechanisms
do not affect large-wavelength, long-time properties, such as growth
exponents and universal scaling functions. In this spirit, Kawasaki dynamics
and the Cahn-Hilliard equation have been extensively used to describe phase
ordering in binary alloys. However, in most real solids, microscopic
atom-atom exchanges can be mediated by a variety of processes \cite{basic MS}%
, with direct exchanges playing a rather small role since steric hindrance
tends to create large energy barriers. The most common mechanisms involve
exchanges with defects, such as vacancies or interstitial sites. For these
reasons, alloys have often been modelled by three-state models \cite
{3statemodel} whose dynamics is controlled by atom-vacancy exchanges, with
direct exchanges being completely forbidden. Two problems in particular have
attracted considerable attention: first, the effect of vacancies on phase
separation and domain growth \cite{VMO,Puri}, and second, their role in
atomic interdiffusion \cite{interdiffusion,alex}. A number of studies have
also addressed vacancy-mediated ordering in a variety of {\em %
antiferromagnetic }alloys \cite{AFM}, as well as surface modes of unstable
droplets in a stable vapor phase \cite{Plapp}.

In this review, we will describe a third aspect of defect-mediated dynamics,
namely, the ``inverse'' of the phase ordering problem. Instead of studying
the growth of order in response to a sudden temperature decrease (a
``quench''), we focus on the {\em disordering} of a finite system, following
a rapid {\em increase} in temperature. Starting with a zero-temperature
ferromagnetic configuration, i.e., a perfectly phase-segregated system with
sharp interfaces, we monitor how the interfaces ``roughen'' and how
particles of one species are transported into regions dominated by the other
species. Clearly, if the final temperature is sufficiently high, the
interfaces will eventually dissolve completely, leaving us with a
homogeneous final state. Several questions emerge quite naturally: Are there
characteristic time scales on which the disordering takes place, and how do
they depend on system size, temperature and other control parameters? How do
local density profiles and correlation functions evolve with time? Are there
any scaling regimes, and what are the appropriate scaling variables? How do
these features respond to changes in the relative concentrations of
vacancies and alloy components?

In the following, we address some of these questions in a simple model for
defect-mediated interface destruction and bulk disordering. We consider a
symmetric (Ising-like) binary alloy of A and B atoms which is diluted by a
very small number of vacancies (defects), reflecting the minute vacancy
concentrations (of the order of $10^{-5}$) found in most real systems.
Following an upquench from {\em zero} to a {\em finite} temperature $T$, the
vacancies act as ``catalysts'' for the disordering process, exchanging with
neighboring particles according to the usual energetics of the (dilute)\
Ising model. The particles themselves form a passive background whose
dynamics is slaved to the defect motion. Thus, this system corresponds to a
real material in which the characteristic time scale for vacancy diffusion
is much faster than the ordinary bulk diffusion time. While vacancies are
typically distributed uniformly in the bulk, certain defects may prefer to
accumulate at the interfaces. Thus, the number of defects is not necessarily
extensive in system size.

While we allow for some variation in the vacancy number, we consider equal
concentrations of A and B atoms. Thus, our work forms a natural complement
to the only other study \cite{ZFP} of vacancy-mediated disordering in the
literature. There, the alloy composition is chosen highly asymmetric: 95\%
of A atoms versus only 5\% for the B species, with a single vacancy. Thus,
the A atoms form a {\em matrix} for a B-{\em precipitate}. The alloy is
first equilibrated at a very low temperature, so that small clusters of B
atoms are present. It is then rapidly heated to a higher temperature, and
the number and size of B clusters are monitored. Three different scenarios
are observed, depending on whether the final temperature is below the
miscibility gap, above the miscibility gap but below $T_{c}$, or above $%
T_{c} $. In the first case, the precipitates remain compact. They dissolve
partially at first, but then equilibrate again by coarsening. In the second
case, the precipitates also remain compact but eventually dissolve
completely, mostly through ``evaporation'' from their surfaces. In the third
case, the clusters decompose rapidly (``explode'') into a large number of
small fragments which then disappear diffusively.

The complete or partial mixing of two materials at an interface plays a key
role in many physical processes, such as corrosion or erosion phenomena \cite
{er+corr}. We mention just two applications with huge technological
potential for device fabrication. The first concerns nanowire etching by
electron beam lithography \cite{Drouin}: If a thin film of platinum is
deposited on a silicon wafer, interdiffusion of Pt and Si produces a mixing
layer. If this layer is heated locally by, e.g., exposure to a conventional
electron beam, silicides, such as Pt$_{2}$Si and PtSi, form. The unexposed
platinum can subsequently be etched away, leaving conducting nanoscale
structures behind. The second example concerns mesoscopic superlattice
structures, consisting of alternating magnetic and nonmagnetic metallic
layers. If adjacent magnetic layers couple antiferromagnetically, the
application of a large uniform magnetic field to these layered structures
results in giant magnetoresistance \cite{GMR}. However, the performance of
these devices requires precisely engineered layer thicknesses and
interfaces, and can be significantly affected by disorder \cite{disorder},
including interdiffusion or interfacial fluctuations.

While motivated by these applications, our study can only form a baseline
here, for further work on more realistic models. However, it also has some
rather fundamental implications. First, it serves as a testing ground for a
basic problem in statistical physics, namely, how a system approaches its
final steady state, starting from an initial {\em non-stationary}
configuration. A second view of our study addresses the effect of a random
walker on its background medium. Each move of a vacancy rearranges the
background atoms slightly, leaving a trail behind like a child running
across a sandy beach.\ In the simplest case, the walker is purely Brownian.
In our language, this corresponds to an upquench to infinite temperature, $%
T=\infty $, where energy barriers are completely irrelevant. In this case,
there is no feedback from the background to the local motion of the vacancy.
Nevertheless, each displaced atom displays its own intriguing dynamics
(mainly in $d=2$) \cite{BH,TZ}. To study the collective behavior of the
atoms, \cite{TKSZ} explored a lattice filled with just two species of
(indistinguishable) particles. Beyond this simple case is a system heated to%
{\em \ finite} temperatures. Since the background affects the vacancy
through the local energetics associated with the next move, no exact
solutions are known. Instead, progress relies mainly on simulations.

The key result of our study is the observation of {\em three} distinct
temporal regimes, separated by two crossover times, provided the final
temperature is not too close to $T_c$. The {\em intermediate and late}
stages of the disordering process exhibit dynamic scaling, with
characteristic exponents and scaling functions that are computed
analytically. For the system sizes considered here, a clear breakdown of
these scaling forms is observed for temperatures within about $10\%$ of $T_c$%
. We argue that this occurs when the correlation length becomes comparable
to the system size. In contrast, the early stage of the disordering process
consists of interfacial destruction in a highly anisotropic manner \cite{TZ}.

This article is organized as follows. We first introduce our model and
define an appropriate ``disorder'' parameter. We then review the purely
diffusive case, corresponding to $T=\infty $ \cite{TKSZ}. Simulation results
for two-dimensiononal systems and an exactly soluble mean-field theory in
general dimension are discussed. In Section 4, we present a number of new
results, concerning upquenches to finite temperatures where
particle-particle interactions come into play. We conclude with some
comments on the effect of external fields, upquenches to temperatures $%
T<T_{c}$, and details concerning interfacial destruction in the early stages.

\section{The Model.}

In this section, we describe the dilute Ising model underlying our Monte
Carlo simulations. It is defined on a two-dimensional ($d=2$) square lattice
of dimension $L\times L$, with sites denoted by a pair of integers, ${\bf %
r\equiv }(x,y)$. The boundary conditions are fully periodic in all
directions. To model the two species of particles, black (``spin up'') and
white (``spin down'', displayed as gray in the figures), and the vacancies,
we introduce a spin variable $\sigma _{{\bf r}}$ at each site which can take
three values: ${+1}$ ($-1$) if the site is occupied by a black (white)
particle, and ${\sigma _{{\bf r}}=0}$ if it is empty. Multiple occupancy is
forbidden. The numbers of black $(N^{+})$ and white $(N^{-})$ particles are
conserved and differ by at most $1$. The number of vacancies, $M$, is much
smaller: $M\ll N^{+}$. In fact, most of our simulations will be restricted
to $M=1$, to model the minute vacancy concentrations in real systems. In our
analytic work, we will also consider a more general case, where $M$ is
allowed to vary with system size according to $M\propto L^{\gamma }$.
Different values of the {\em vacancy number }exponent{\em \ }$\gamma \in
[0,d]$ will be discussed. Clearly, the single vacancy case corresponds to $%
\gamma =0$.

The particles and vacancies interact with one another according to a dilute
Ising model: 
\begin{equation}
{\cal H}[\sigma _{{\bf r}}]=-J\sum_{\left\langle {\bf r},{\bf r}^{\prime
}\right\rangle }\sigma _{{\bf r}}\sigma _{{\bf r}^{\prime }}  \label{Ham}
\end{equation}
with a ferromagnetic, nearest-neighbor coupling $J>0$. Since the dilution is
so small, the behavior of the system is that of the ordinary (non-dilute)
two-dimensional Ising model. Thus, it has a phase transition, from a
disordered to a phase-segregated phase, at the Onsager critical temperature, 
$T_{c}=2.267...J/k_{B}$ \cite{Onsager}. The ground state is doubly
degenerate. It consists of a strip of black and a strip of white particles,
each filling half the system, separated by two planar interfaces running
parallel to a lattice axis. Since the particle-particle interactions are
ferromagnetic, the few vacancies accumulate at the interfaces.

Next, we turn to the dynamics of the model. Only particle-hole exchanges are
allowed and performed with the usual Metropolis \cite{Metropolis} rate: $%
\min \{1,\exp (-\beta \Delta {\cal H})\}$, where $\Delta {\cal H}$ is the
energy difference of the system before and after the jump, and $\beta
=1/(k_BT)$ is the inverse temperature. The initial configuration of the
system is perfectly phase-segregated, with two interfaces chosen to lie
along the $x$-axis. The vacancies are located at random positions along one
of the interfaces. Since this is technically a zero-temperature
configuration, while the dynamics occurs at temperature $T>0$, the vacancies
will move around, disordering the interfaces and, if $T>T_c$, dissolving
them eventually. The final steady state is an equilibrium state of the usual
two-dimensional Ising model. Therefore, many of its properties are exactly
known \cite{MW}.

In our simulations, the system sizes range from $L=30$ to $60$. The final
temperature $T$, measured in units of the Onsager temperature $T_{c}$,
varied between $1.1T_{c}$ and infinity. Our data are averages over $10^{2}$
to $10^{4}$ realizations (or runs) depending on the desired quality of the
data. The time unit is one Monte Carlo step (MCS) which corresponds to $M$
attempted particle-hole exchanges. All systems investigated equilibrate
after about $10^{8}$ MCS.

To monitor the evolution of the system, we measure a ``disorder parameter'',
defined as the average number of black and white nearest-neighbor pairs, $%
{\cal A}(L,t)$, as a function of (Monte Carlo) time $t$. This quantity is
easily related to the Ising energy, 
\begin{equation}
{\cal A}(L,t)=\frac{d}{2}L^{d}+\frac{1}{2J}\left\langle {\cal H}%
\right\rangle +O(L^{\gamma })  \label{DP}
\end{equation}
where $\left\langle \cdot \right\rangle $ denotes the configurational
average over runs. The correction is due to the vacancies and remains much
smaller than the two leading terms. More detailed information is carried by
the local hole and magnetization densities, defined respectively as 
\begin{eqnarray}
\phi ({\bf r},t) &\equiv &\left\langle \delta _{{\sigma _{{\bf r}},}%
0}\right\rangle \text{ }  \nonumber \\
\psi ({\bf r},t) &\equiv &\left\langle \sigma _{{\bf r}}\right\rangle \text{ 
}  \label{dens}
\end{eqnarray}
The Kronecker-$\delta $ ensures that lattice site ${\bf r}$ is occupied by
the vacancy. Non-zero values of $\psi ({\bf r},t)$ indicate an {\em excess}
of white or black particles at lattice site ${\bf r}$, which is obviously a
sensitive measure of the disordering process. The full time dependence of
these densities can in general only be computed within a mean-field
approach. However, their stationary forms are easily found from exactly
known properties of the two-dimensional Ising model.

\section{Brownian Vacancies:\ $T=\infty $.}

In this Section, we focus on the simplest case, namely, upquenches to
infinite temperature. As a consequence, the nearest-neighbor coupling $J$
plays no role at all, and all attempted particle-hole exchanges are executed
with unit rate. Thus, the vacancies perform a Brownian random walk,
regardless of their local environment. We first summarize our simulation
results which suggest the key to the mean-field analysis, namely, a
separation of time scales. We then introduce a suitable set of exponents and
scaling forms and show that the late stages of the disordering process
display dynamic scaling. Finally, we turn to a mean-field theory and compute
these exponents and scaling functions analytically.

\subsection{Simulation Results}

\smallskip To illustrate the gradual destruction of the interfaces and the
disordering of the bulk, Fig.~1 shows the evolution of a typical
configuration, in a $60\times 60$ system with a single vacancy (which is
represented by a white square in the figures). At $t=0$, the vacancy is
located at the interface in the center which begins to break up slowly.
Eventually, the second interface also becomes affected, as more and more
particles are transported into regions of opposite color, until the system
finally disorders completely. For later reference, we note that the {\em %
last two} configurations, at $10^{7}$ and $10^{8}$ MCS, are both already
fully random. The disordering process is clearly reflected in the number of
broken bonds: ${\cal A}(60,t)$, shown in Fig.~2, increases from its minimum
of $O(L)$ for the initial configuration at $t=0$, to $O(L^{2})$ for the
fully equilibrated system at $t=10^{7}$. One clearly distinguishes three
regimes, shown schematically in the inset: an {\em early }regime (I), the 
{\em intermediate}, or {\em scaling}, regime (II), and finally a {\em late }%
or{\em \ saturation }regime (III) in which the system has effectively
reached the steady state. Tracking the motion of the vacancy, the physical
origin of these three regimes is easily identified. For {\em early} times
(regime I), the vacancy is still localized in the vicinity of its starting
point, far from the boundaries of the system. After a time of $O(L^{2})$,
however, the vacancy has explored the whole system and is effectively
equilibrated. This marks the onset of the {\em intermediate} regime. The
particle distribution is still strongly inhomogeneous here and does not
equilibrate until the system enters the {\em saturation} regime. We
emphasize that the second and third regimes emerge only in a {\em finite}
system. In an infinite system, regime I persists for all times. Some aspects
will be briefly summarized in the concluding remarks.

\begin{figure}[tbp]
\input{epsf}
  \hfill\hfill
\begin{minipage}{0.2\textwidth}
  \epsfxsize = \textwidth \epsfysize = 1\textwidth \hfill
  \epsfbox{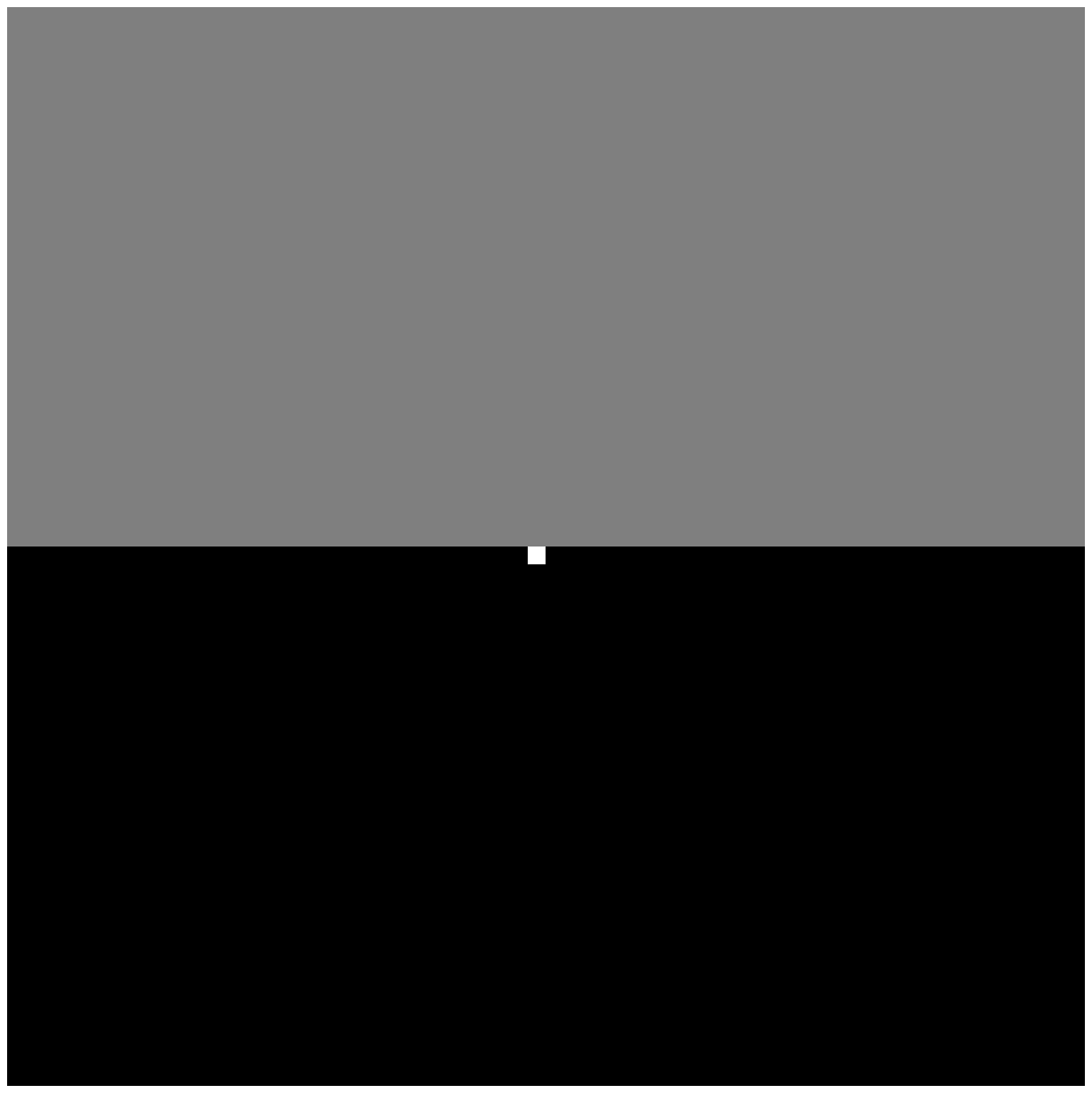} \hfill
    \begin{center} \large{$0$  MCS} \end{center}
  \end{minipage}
  \hfill \hfill
\begin{minipage}{0.2\textwidth}
  \epsfxsize = \textwidth \epsfysize = 1\textwidth \hfill
  \epsfbox{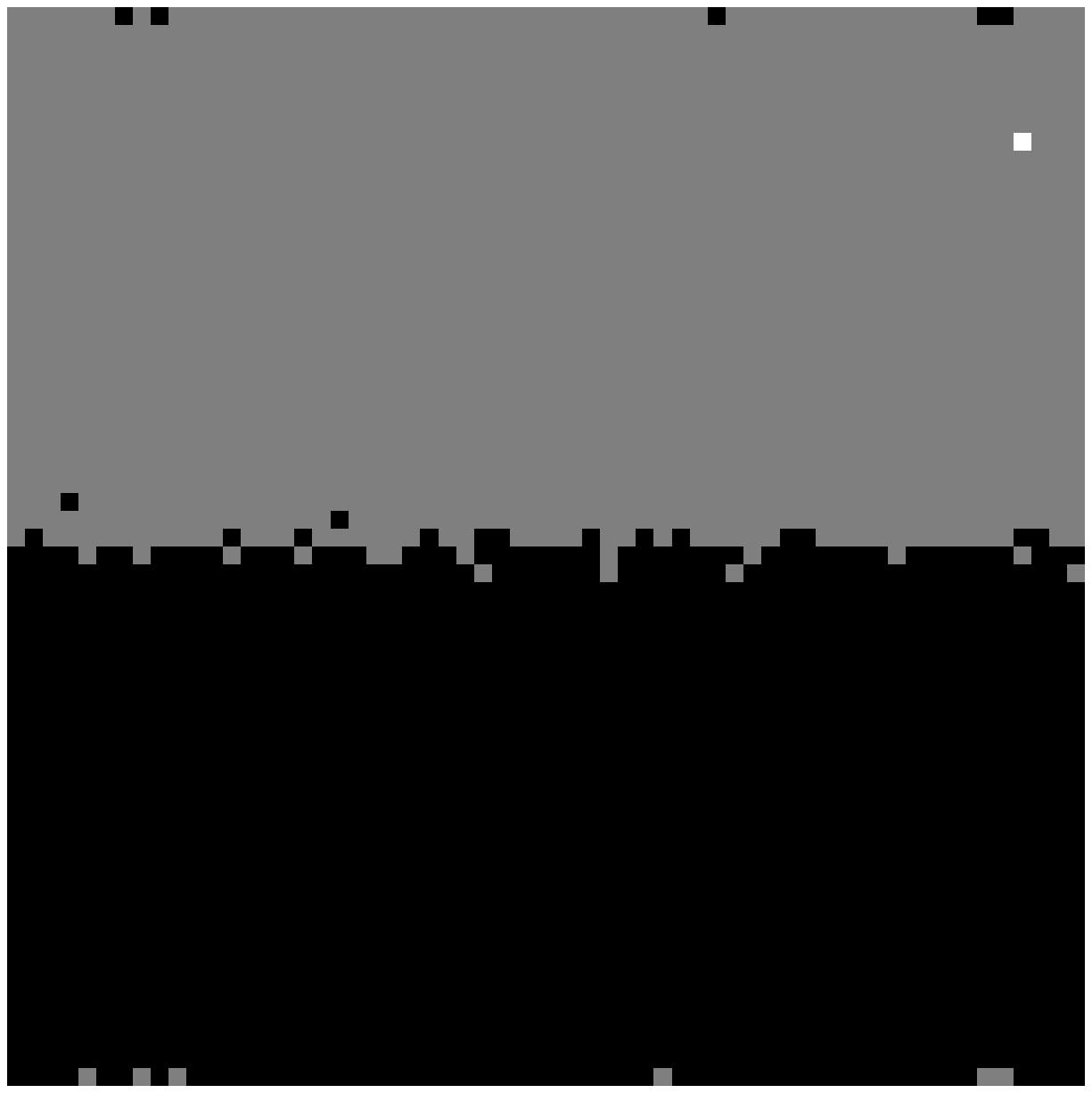} \hfill
    \begin{center} \large{$10^{4}$ MCS} \end{center}
  \end{minipage}
  \hfill \hfill
\begin{minipage}{0.2\textwidth}
  \epsfxsize = \textwidth \epsfysize = 1\textwidth \hfill
  \epsfbox{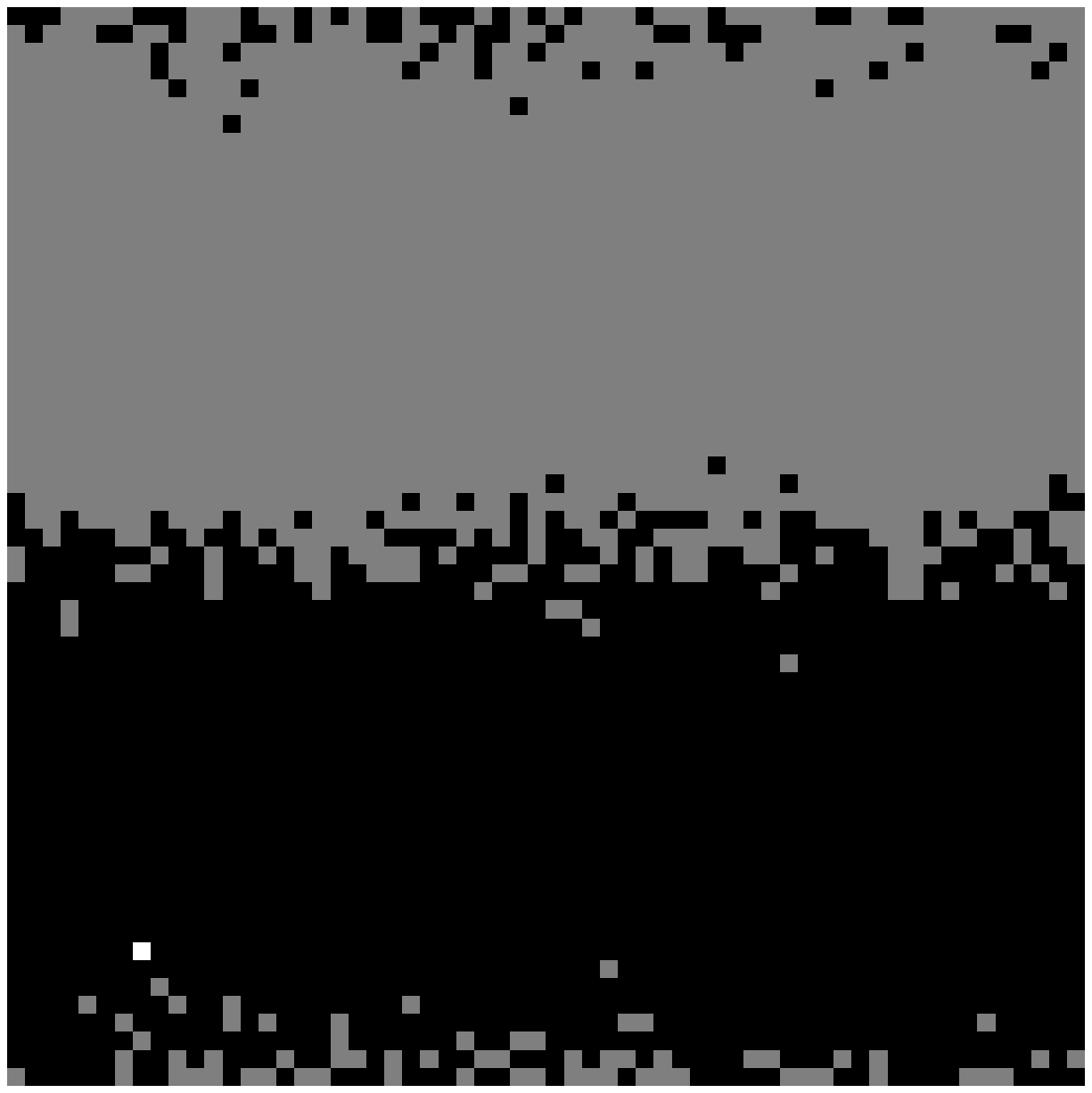} \hfill
    \begin{center} \large{ $10^{5}$ MCS} \end{center}
  \end{minipage}
  \hfill\hfill \vspace{-0.02\textwidth}
\par
\hfill\hfill
\begin{minipage}{0.2\textwidth}
  \epsfxsize = \textwidth \epsfysize = 1\textwidth \hfill
  \epsfbox{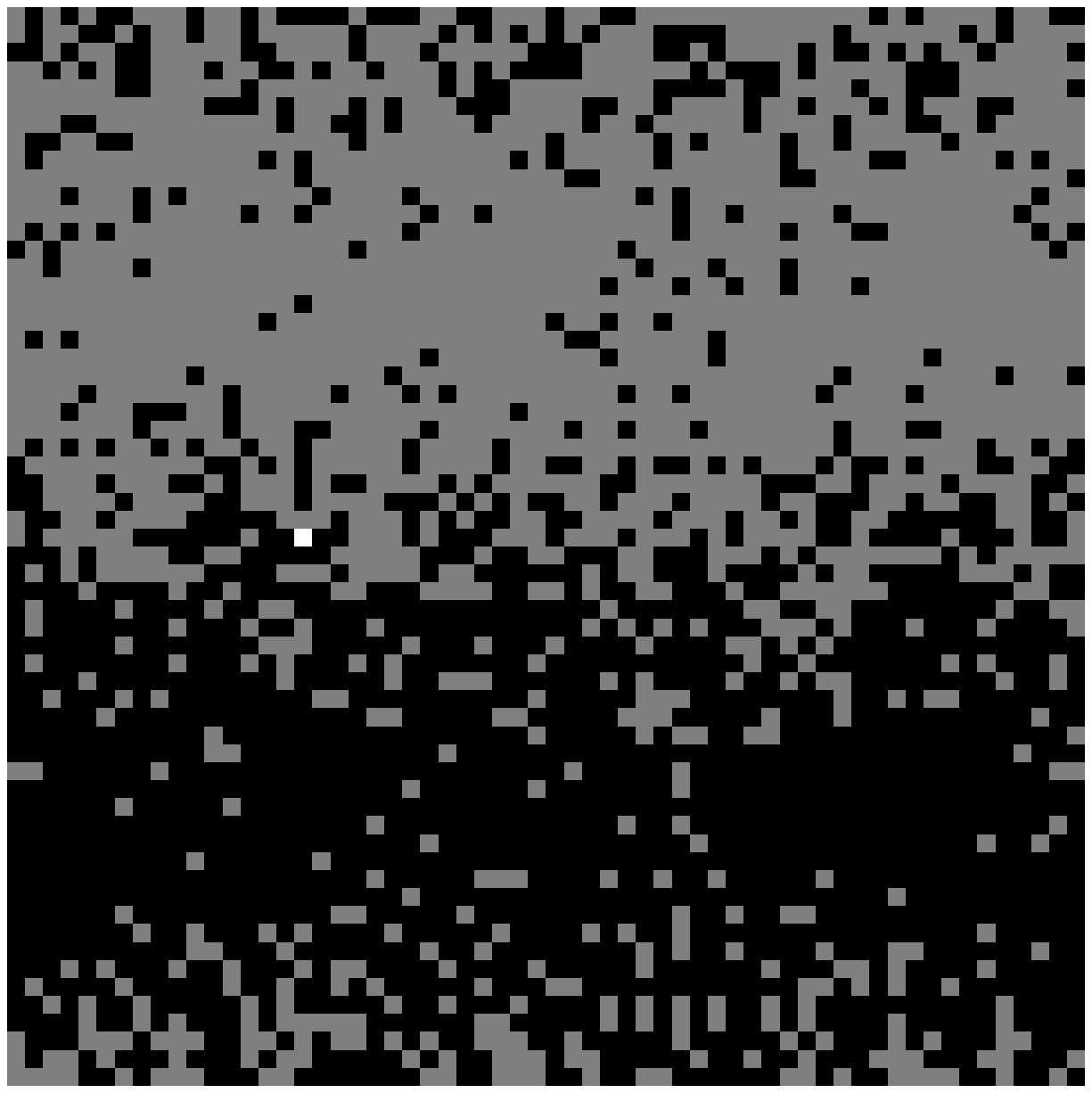}
    \begin{center} \large{$10^{6}$ MCS}  \end{center}
  \end{minipage}
  \hfill \hfill
\begin{minipage}{0.2\textwidth}
  \epsfxsize = \textwidth \epsfysize = 1\textwidth \hfill
  \epsfbox{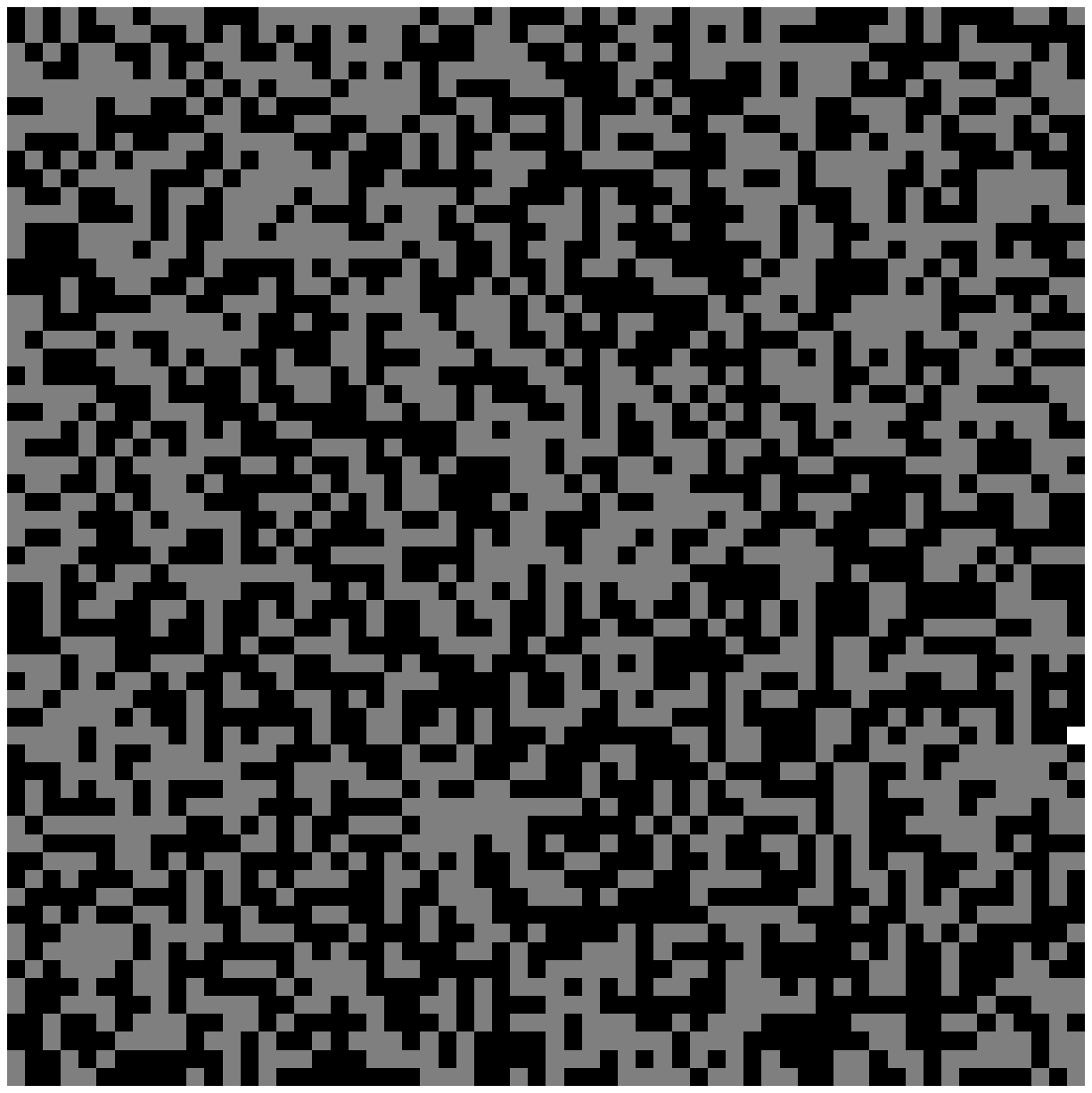}
    \begin{center} \large{$10^{7}$ MCS} \end{center}
  \end{minipage}
  \hfill \hfill
\begin{minipage}{0.2\textwidth}
  \epsfxsize = \textwidth \epsfysize = 1\textwidth \hfill
  \epsfbox{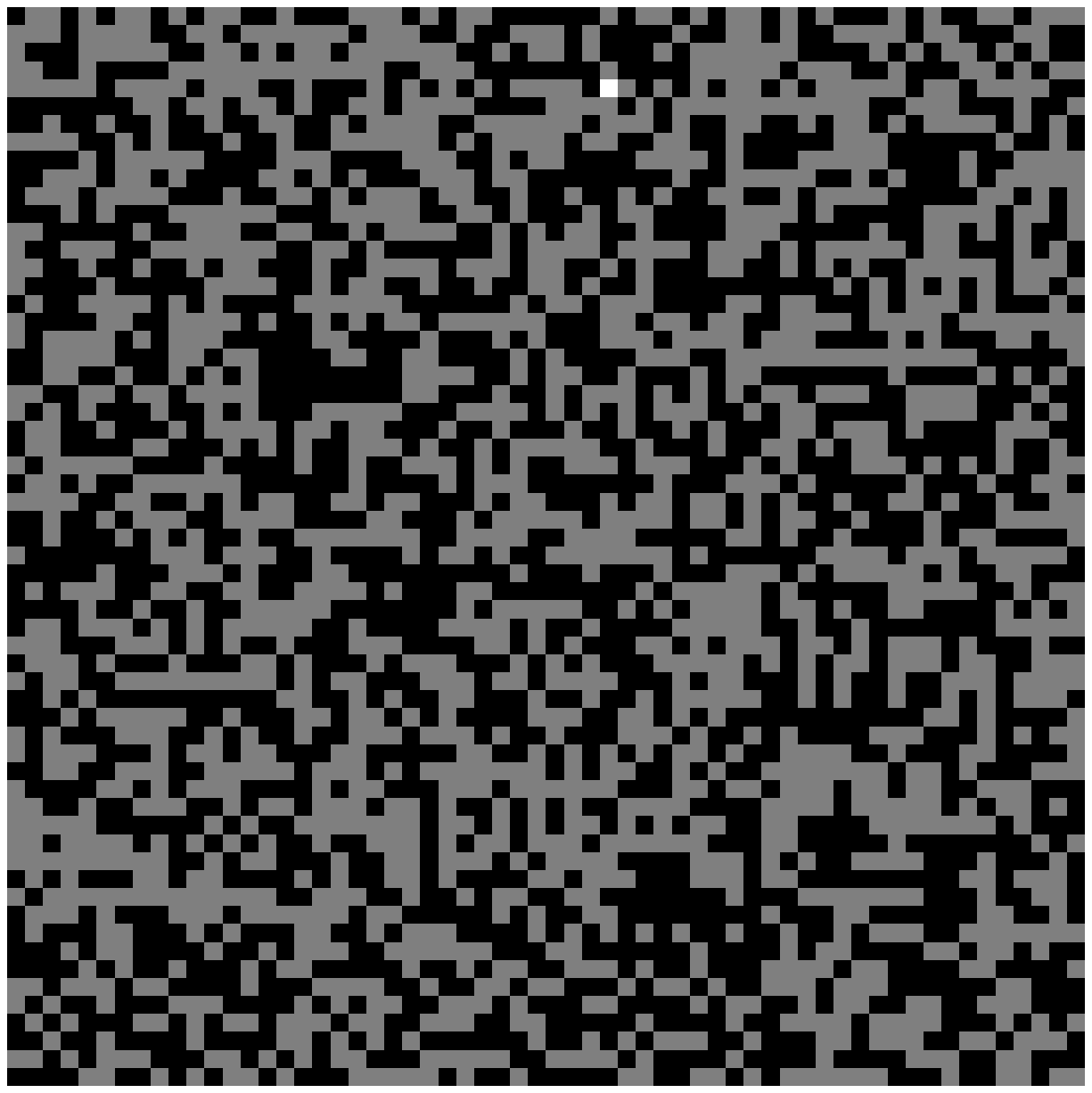}
    \begin{center} \large{ $10^{8}$ MCS} \end{center}
  \end{minipage}
  \hfill \hfill
\vspace{0.03\textwidth}
\caption{Sequence of snapshots showing the disordering process of a 
$60 \times 60$
system with $T= \infty$. The black and gray squares represent the 
two types of particles($\sigma = \pm 1$) and the white square denotes the
vacancy($\sigma = 0$).  The configurations were recorded after  $0, 10^4,
 10^5, 10^6, 10^7 \ $and$ \ 10^8$ MCS.  }
\label{6latticepic}
\end{figure}
\noindent
\begin{figure}[tbp]
\input{epsf}
\begin{center}
\begin{minipage}{0.5\textwidth}
  \epsfxsize = \textwidth \epsfysize = 1.0\textwidth \hfill
  \epsfbox{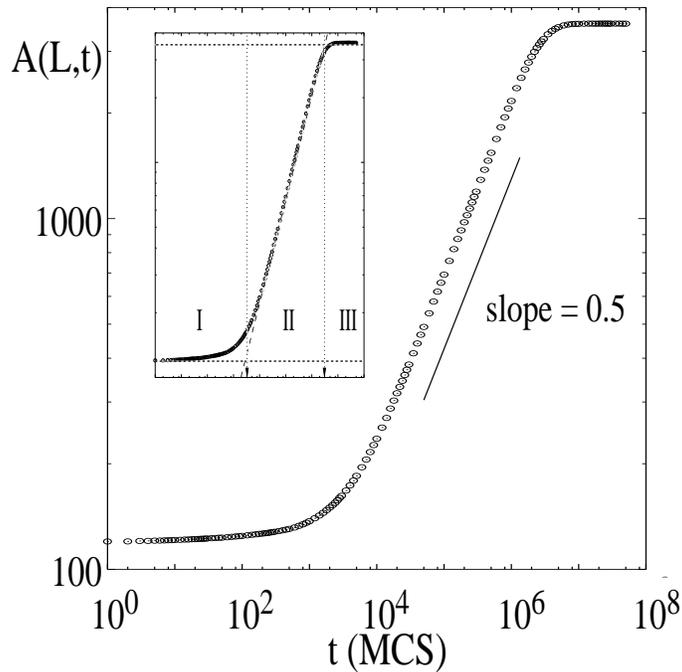}
    \vspace{-.2cm}
\end{minipage}
\end{center}
\caption{Plot of the total number of broken bonds, $A(L,t)$ vs  
$t$, for $L=60$ and $T=\infty$. It  
shows the emergence  of an {\em early }regime (I), an  
{\em intermediate}, or {\em scaling}, regime (II), and a {\em late }or
{\em saturation }regime (III). The straight reference line has slope 
$0.5$.}
\label{p1b}
\end{figure}

Our key observation is that, independent of dimension, Regimes II and III
exhibit {\em dynamic scaling}. To characterize this behavior, we introduce a
set of exponents: First, the saturation value of ${\cal A}$ scales with
system size according to $\lim_{t\rightarrow \infty }{\cal A}(L,t)\equiv 
{\cal A}_{sat}(L)\sim L^{\alpha }$. Second, in the intermediate regime, $%
{\cal A}(L,t)$ grows as ${\cal A}(L,t)\sim L^{\sigma }t^{\beta }$. Finally,
the two crossover times (``early'' and ``late'') scale as $t_{e}\sim
L^{z_{e}}$, and $t_{l}\sim L^{z_{l}}$. Thus, the intermediate-to-late time
crossover can be summarized by the scaling form 
\begin{equation}
{\cal A}(L,t)\sim L^{\alpha }f\left( t/t_{l}\right)  \label{Asc}
\end{equation}
with a scaling function $f$ which satisfies $f(x)\simeq const$ for $x\gg 1$,
and $f(x)\sim x^{\beta }$ for $x\ll 1$ (but large enough to fall within
Regime II). The consistency condition ${\cal A}(L,t_{l})\simeq {\cal A}%
_{sat}(L)$ yields the scaling law 
\begin{equation}
\sigma +z_{l}\beta =\alpha .  \label{scl}
\end{equation}
Due to the presence of the novel exponent $\sigma $, (\ref{scl}) {\em %
differs }from the familiar $z\beta =\alpha $ which controls surface growth
in, e.g., the Edwards-Wilkinson \cite{EW} or KPZ \cite{KPZ} models.

We first consider a single vacancy, in $d=2$. For this case, the data give $%
\alpha =2\pm 0.1$, $\beta =0.5\pm 0.06$, $z_{e}=2\pm 0.2$, $z_{l}=4\pm 0.2$,
and $\sigma =0\pm 0.1$. Clearly, these exponents satisfy the scaling law (%
\ref{scl}). Excellent data collapse is obtained by plotting ${\cal A}%
(L,t)/L^{\alpha }$ versus $t/L^{z_{l}}$, shown in Fig.~3 for a range of
system sizes. The same indices are observed for several vacancies, provided
their number remains {\em constant} as the system size $L$ is varied,
corresponding to a vacancy exponent $\gamma =0$. We note briefly that the
exponent $\sigma $ is required here to accomplish good data collapse in
dimensions greater than $2$.

Another natural case is $\gamma =d-1$, corresponding to a situation where
the defects are initially ``frozen'' at the interfaces, and their number
scales with the interfacial area. Such a scenario could occur in surface
catalysis or in welding. Since our data are restricted to $d=2$, we examine
the case $\gamma =1$, i.e., $M\propto L$. Here, the data collapse only if 
{\em (i) }curves corresponding to different system sizes are shifted by an $%
M $-dependent factor along the $\ln t$-axis, and {\em (ii) }the late
crossover exponent takes on the new value $z_{l}=3\pm 0.2$, while $\alpha $
and $\beta $ remain unchanged. This behavior suggests that the exponent $%
\sigma $ depends {\em non-trivially} on $\gamma $. For $\gamma =1$, the data
are consistent with $\sigma =1/2$, so that the scaling form (\ref{Asc})
still holds and the scaling law (\ref{scl}) is satisfied with the modified
exponent $z_{l}$.

This concludes the discussion of our simulation results. We now turn to an
analytic description of the disordering process.

\begin{figure}[tbp]
\input epsf
\begin{center}
\begin{minipage}{.5\textwidth}
  \epsfxsize = \textwidth \epsfysize = 1.0\textwidth \hfill
  \epsfbox{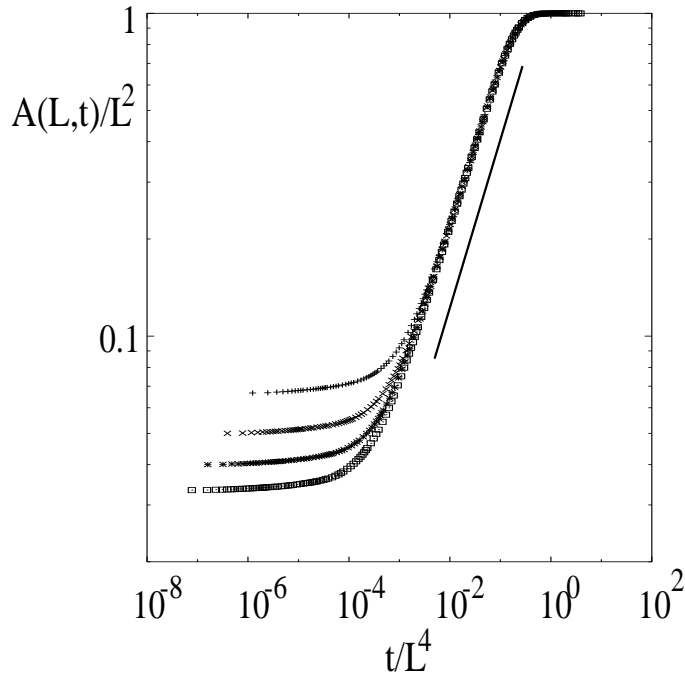}
    \vspace{-.2cm}
\end{minipage}
\end{center}
\caption{The scaling plot of $A(L,t)/L^{2}$ vs 
$t/L^{4}$ for $L=30-60$ with $T= \infty$. The straight reference line
has slope $0.5$.}
\label{p2}
\end{figure}
\subsection{Mean-Field Theory}

As a first step, we summarize the exactly known results for the initial
configuration and the final, fully disordered state. The analysis is easily
performed in general dimension, for a hypercube of side length $L$ with
fully periodic boundary conditions. The initial value of the disorder
parameter is given by the number of broken bonds across the two initial,
flat interfaces between black and white particles: ${\cal A}%
(L,t=0)=2L^{d-1}+O(M)$. The saturation value, ${\cal A}_{sat}(L)=\frac d2%
L^{2d}/(L^d-1)$ follows easily from the fact that the final steady state is
completely random for the Brownian vacancy case with $T=\infty $. Thus, we
can read off the {\em exact} result $\alpha =d$.

To proceed further, we derive a set of equations of motion, for the
coarse-grained local hole and particle densities. Starting from the
microscopic master equation, we can easily obtain evolution equations for
the discrete densities, $\phi ({\bf r},t)$ and $\psi ({\bf r},t)$. Since
there is no feedback from the particle background, the equation for $\phi (%
{\bf r},t)$ is completely independent of $\psi ({\bf r},t)$. In contrast,
the motion of the particles is slaved to that of the vacancy, so that the
equation for $\psi ({\bf r},t)$ is inherently nonlinear and contains
two-point functions of the form $\left\langle \sigma _{{\bf r}^{\prime
}}\delta _{{\sigma _{{\bf r}},}0}\right\rangle $. Here, a mean-field
assumption is required, to truncate these averages. Finally, we take the
continuum limit, by letting the lattice constant vanish at fixed system size 
$L$. For simplicity, we use the same notation for discrete densities and
their continuous counterparts. The position vector ${\bf r}\equiv
(x_{1,}....,x_d)$ denotes a point in the hypercube $-L/2\leq x_i\leq +L/2$, $%
i=1,2,...,d$, with volume $V\equiv L^d$. We also retain the symbol $t$ for
time, since Monte Carlo time and its coarse-grained counterpart differ only
by a scale factor. After some suitable rescalings, we obtain the desired
mean-field equations: 
\begin{eqnarray}
\partial _t\phi ({\bf r},t) &=&\nabla ^2\phi ({\bf r},t)  \nonumber \\
\partial _t\psi ({\bf r},t) &=&\phi ({\bf r},t)\nabla ^2\psi ({\bf r}%
,t)-\psi ({\bf r},t)\nabla ^2\phi ({\bf r},t).  \label{EoMs}
\end{eqnarray}
The simple diffusion equation for $\phi $ reflects the Brownian random walk
of the vacancies. The magnetization density $\psi $ obeys a balance
equation: the first term reflects a gain, provided a vacancy is initially
present and a particle ``diffuses'' in from a neighboring site. The second
term accounts for a loss, due to a particle jumping to a vacant
nearest-neighbor site. Similar equations have been discussed in the context
of biased diffusion of two species \cite{hsz} and vacancy-mediated
interdiffusion \cite{alex}. Note that both equations take the form of
continuity equations, due to the conservation laws on the particle numbers.
The two densities are normalized, according to $\int_V\phi ({\bf r},t)=M$,
and $\int_V\psi ({\bf r},t)=0$. Given fully periodic boundary conditions,
these equations have to be supplemented with appropriate initial conditions.
For our simulations, we choose $\phi ({\bf r},0)=\frac M{L^{d-1}}\delta (y)$
where $y\equiv x_d$, and $\psi ({\bf r},0)=2\theta (y)-1$. The final state
is, of course, trivial: $\phi ({\bf r},\infty )\equiv \phi _o=M/V$, and $%
\psi ({\bf r},\infty )=0$.

To solve for the hole and magnetization densities, we seek the separation of
time scales, first observed in the Monte Carlo data, in the equations (\ref
{EoMs}). The hole density, with a diffusion coefficient of $O(1)$, relaxes
rapidly to the final value $M/V$. By contrast, the magnetization density is
essentially slaved to $\phi ({\bf r},t)$ and relaxes with a diffusion
``coefficient'' of $O(1/V)$. The hole density spreads diffusively and
reaches the boundaries of the system after a time of $O(L^{2})$, marking the
end of the early regime. Thus, we identify the {\em early } crossover time $%
t_{e}\propto L^{2}$ and read off $z_{e}$ $=2$. For later times, the
vacancies are uniformly distributed over the system, so that $\phi $ may be
replaced by its stationary value, $\phi _{o}=\frac{M}{V}$. Inserting this
into the equation for $\psi $ results in a simple diffusion equation 
\begin{equation}
\partial _{t}\psi =\phi _{o}\nabla ^{2}\psi  \label{psi}
\end{equation}
with a diffusion coefficient $M/V$. Its solution, subject to the initial and
fully periodic boundary conditions, is easily found: 
\begin{equation}
\psi ({\bf r},t)=\frac{4}{\pi }\sum_{n=1}^{\infty }\frac{sin[2\pi (2n-1)y/L]%
}{2n-1}\;e^{-\epsilon \,t(2n-1)^{2}}\;,  \label{psisol}
\end{equation}
where $\epsilon \equiv 4\pi ^{2}\phi _{o}/L^{2}$. Of course, $\psi ({\bf r}%
,t)$ depends on $y$ only, reflecting the homogeneity of initial and boundary
conditions in the $d-1$ transverse directions. We also note that, while this
form is exact, it converges rapidly only for late times, $\epsilon \,t\gg 1$%
, corresponding to the saturation regime. There, the profile is harmonic
with a rapidly decaying amplitude: $\psi ({\bf r},t)\simeq \frac{4}{\pi }%
sin[2\pi y/L]e^{-\epsilon \,t}$. Below, we will present an equivalent form
with good convergence in the opposite limit.

To describe the simulation results, we need an expression for the disorder
parameter ${\cal A}$. Recalling its connection to the average Ising energy,
Eqn (\ref{DP}), we invoke the continuum limit of the latter \cite{Amit}, $%
\left\langle {\cal H}\right\rangle =-dJ\int_{V}\psi ({\bf r},t)^{2}$, which
is correct up to surface terms of $O(1/L^{d-1})$. Thus, we obtain 
\begin{equation}
{\cal A}(L,t)=\frac{d}{2}\left[ V-\int_{V}\psi ({\bf r},t)^{2}\right] \quad .
\label{Acl}
\end{equation}

We note that this form is manifestly extensive. Moreover, since surface
terms have been dropped, the initial value of ${\cal A}$ is now simply $0$.

The time evolution of this quantity follows from (\ref{psisol}): 
\begin{equation}
{\cal A}(L,t)=\frac{d}{2}V\left[ 1-S(2\epsilon t)\right] \;,  \label{Acl1}
\end{equation}
where $S(\zeta )\equiv \frac{8}{\pi ^{2}}\sum_{1}^{\infty }e^{-\zeta
(2n-1)^{2}}/(2n-1)^{2}$. Since $S(0)=1$ and $S(\infty )=0$, we verify that $%
{\cal A}$ does take on the correct initial and final values. It is also
consistent with our postulated scaling form, Eqn (\ref{Asc}), provided we
identify the scaling exponent $\alpha $ with $d$ and the scaling variable $%
t/t_{l}$ with $\epsilon t$. This suggests to define the late crossover time
as $t_{l}\equiv 1/\epsilon $. Collecting the dependence on system size, we
identify $t_{l}\propto L^{2}/\phi _{o}=L^{d+2-\gamma }$. Hence, we read off $%
z_{l}=d+2-\gamma $ for the scaling exponent which controls the late
crossover time. Thus, $\epsilon t\gg 1$ marks the saturation regime where
Eqn (\ref{Asc}) converges well, exhibiting a simple exponential approach to
the saturation value: ${\cal A}(L,t)=\frac{d}{2}V\left[ 1-\frac{8}{\pi ^{2}}%
e^{-2\epsilon t}+O(e^{-4\epsilon t})\right] $.

In contrast, the intermediate regime (II) corresponds to $\epsilon t\ll 1$.
To capture the time dependence here, we re-express the infinite sum via a
Poisson resummation \cite{Jones}. The key advantage of this procedure is to
convert a sum with rapid convergence in one limit ($\epsilon t\gg 1$) into
an equivalent sum with good convergence in the opposite ($\epsilon t\ll 1$)
limit. Deferring technical details to an Appendix, we introduce $u_m\equiv
\pi m/\sqrt{8\epsilon t}$ and arrive at 
\begin{equation}
{\cal A}(L,t)\simeq \frac{2d}{\pi ^{3/2}}V\sqrt{2\epsilon t}\left\{
1+2\sum_{m=1}^\infty (-1)^m\left[ e^{-u_m^2}-u_m\Gamma \left( \frac 12%
,u_m^2\right) \right] \right\} \;  \label{contlim3}
\end{equation}
where $\Gamma (\bullet ,\bullet )$ denotes the incomplete Gamma function. In
this form, the sum over $m$ is suppressed for small $\epsilon t$. Thus, $%
{\cal A}(L,t)\propto V\sqrt{2\epsilon t}\propto L^d\sqrt{Mt/L^{2+d}}$,
yielding the remaining indices, namely $\beta =\frac 12$ independent of
dimension, and $\sigma =\frac 12(d+\gamma -2)$. It is straightforward to
check that our exponents satisfy the scaling relation (\ref{scl}). In two
dimensions, our theory predicts $z_l=4$ and $\sigma =0$ for $\gamma =0$,
while $z_l=3$ and $\sigma =\frac 12$ for $\gamma =1$, in complete agreement
with the Monte Carlo data.

Since the full magnetization profile contains more detailed information than
the disorder parameter, one might also be interested in its form for early
times, $\epsilon t\ll 1$. A similar Poisson resummation, of Eqn (\ref{psisol}%
), results in 
\begin{equation}
\psi ({\bf r},t)=%
\mathop{\rm erf}%
\left( \frac{\tilde{y}}{2\sqrt{\epsilon t}}\right) +\sum_{m=1}^{\infty
}(-1)^{m}\left[ 
\mathop{\rm erf}%
\left( \frac{\tilde{y}+\pi m}{2\sqrt{\epsilon t}}\right) +%
\mathop{\rm erf}%
\left( \frac{\tilde{y}-\pi m}{2\sqrt{\epsilon t}}\right) \right] \;,
\label{psisolP}
\end{equation}
where $\tilde{y}\equiv 2\pi y/L$. For $\epsilon t\ll 1$, this form still
reflects the sharpness of the initial $\Theta $-function which is only
gradually ``washed out''.

This concludes our discussion of the mean-field theory, for the Brownian
vacancy case. We note briefly that our results remain unchanged (up to a
trivial amplitude) if periodic boundary conditions are replaced by
reflecting ones. We should also reiterate that our focus here rests on the
intermediate and late regimes. A full microscopic treatment of the early
regime is beyond the scope of this paper. For the benefit of the reader,
however, a short summary is provided in the conclusions. Instead, we turn to
the role of interactions.

\section{The Effect of Interactions: $T_{c}<T<\infty $.}

When the final temperature of the system is {\em finite}, $T<\infty $, the
full Hamiltonian, Eqn (\ref{Ham}), comes into play when a vacancy attempts
to move. In particular, the defects no longer perform a simple random walk,
since the jump rates now carry information about the local environment,
i.e., the distribution of black and white particles around the originating
and receiving site. Thus, a feedback loop between vacancy and background is
established, in stark contrast to the case $T=\infty $. One might expect
that this has significant consequences for profiles and disorder parameter,
and that dynamic scaling exponents might be modified. It is immediately
obvious that we should distinguish final temperatures above criticality from
those below. In the former case, the final steady-state configurations are
still homogeneous, even if correlations become more noticeable. In the
latter case, however, the equilibrium state itself is phase-separated, so
that the only effect of the vacancies is to ``soften'' the initial
interfaces. This aspect of vacancy-induced disordering should be particulary
interesting in two dimensions, where Ising interfaces are known to be rough 
\cite{RT}. There, the vacancies must mediate both the development of an
``intrinsic interfacial width'' and the emergence of large-scale wanderings
of the interface \cite{interfacial width}. Certainly, we expect these two
phenomena to take place on drastically different time scales, since they are
associated with local {\em vs. }global disorder.

In this Section, we focus on final temperatures {\em above} criticality: $%
T_{c}<T<\infty $, and a single vacancy, $M=1$. Starting with a set of
typical evolution pictures, we investigate the dynamic scaling properties of
the disorder parameter. Since the equations of motion now become highly
nonlinear, we have not been able to solve them exactly. Thus, we offer only
a few comments in conclusion.

Fig.~4 shows the evolution of a typical configuration, in a $60\times 60$
system with a single vacancy at $T=1.5T_{c}$. Note that the MC times are the
same as in Fig.~1. Again, we observe the gradual, yet eventually complete,
destruction of the interfaces. In comparison with Fig.~1, however, two
obvious differences emerge: First, the system takes longer to reach the
final stationary state. Second, the final configuration shows clear evidence
of a $finite$ correlation length. Both features are induced by the
interactions. In particular, the disordering process is slowed down since
the breaking of bonds is energetically costly. At a more quantitative level,
this is documented by Fig.~5 which shows the disorder parameter, for a $%
60\times 60$ system and several temperatures. One observes that the late
crossover time shifts to later times as the temperature decreases. In
contrast, the early crossover time appears to be less affected. Also, the
saturation value of the disorder parameter decreases. However, being
essentially the Ising energy, it remains extensive, with a $T$-dependent
amplitude.

It is natural to ask whether the scaling forms found in the previous section
still hold. In Fig.~6, we present the scaling plots for the disorder
parameter, i.e., we show ${\cal A}(L,t)/L^{d}$ plotted vs. $t/L^{4}$. Each
graph corresponds to a different temperature: $T=3.5T_{c}$ (Fig.~6a), $%
T=1.5T_{c}$ (Fig.~6b), and $T=1.1T_{c}$ (Fig.~6c). Focusing on Fig.~6a, it
is quite evident that the data collapse just as well for $T=3.5T_{c}$ as for 
$T=\infty $ (cf. Fig.~3). The three temporal regimes are well separated and
easy to distinguish. In the intermediate regime, the disorder parameter
follows the $t^{1/2}$ power law, over more than three decades. Thus, the
data at this temperature still obey the scaling form (\ref{Asc}), with the
same set of exponents. However, for the next temperature, $T=1.5T_{c}$ 
(Fig.~6b), the range of perfect data collapse begins to shrink: 
the curve for the
smallest system size ($L=30$) merges noticeably later than in Fig.~6a. Also,
the power law no longer persists over such a wide region, and the exponent $%
\beta $ becomes more difficult to determine. Still, the larger system sizes
collapse rather well, so that there is still a well-defined scaling function
even if its form appears to change. Finally, Fig.~6c shows the data for $%
T=1.1T_{c}$: here, the imposed scaling form manifestly fails to match the
data.

\vspace{-0.08\textwidth}
\begin{figure}[tbp]
 \input epsf
  \hfill\hfill
\begin{minipage}{0.2\textwidth}
  \epsfxsize = \textwidth \epsfysize = 1\textwidth \hfill
  \epsfbox{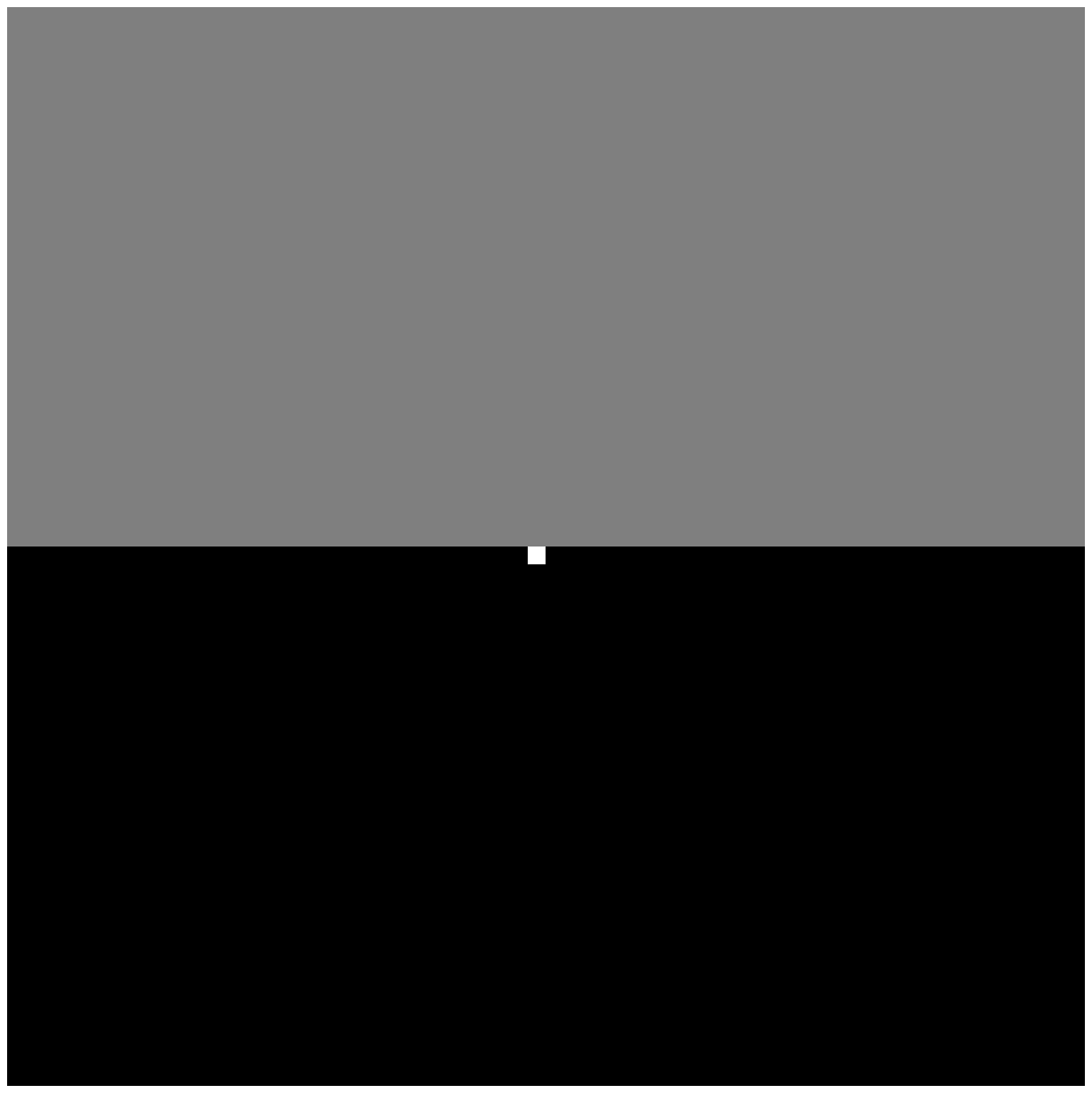} \hfill
    \begin{center} \large{$0$  MCS} \end{center}
  \end{minipage}
  \hfill \hfill
\begin{minipage}{0.2\textwidth}
  \epsfxsize = \textwidth \epsfysize = 1\textwidth \hfill
  \epsfbox{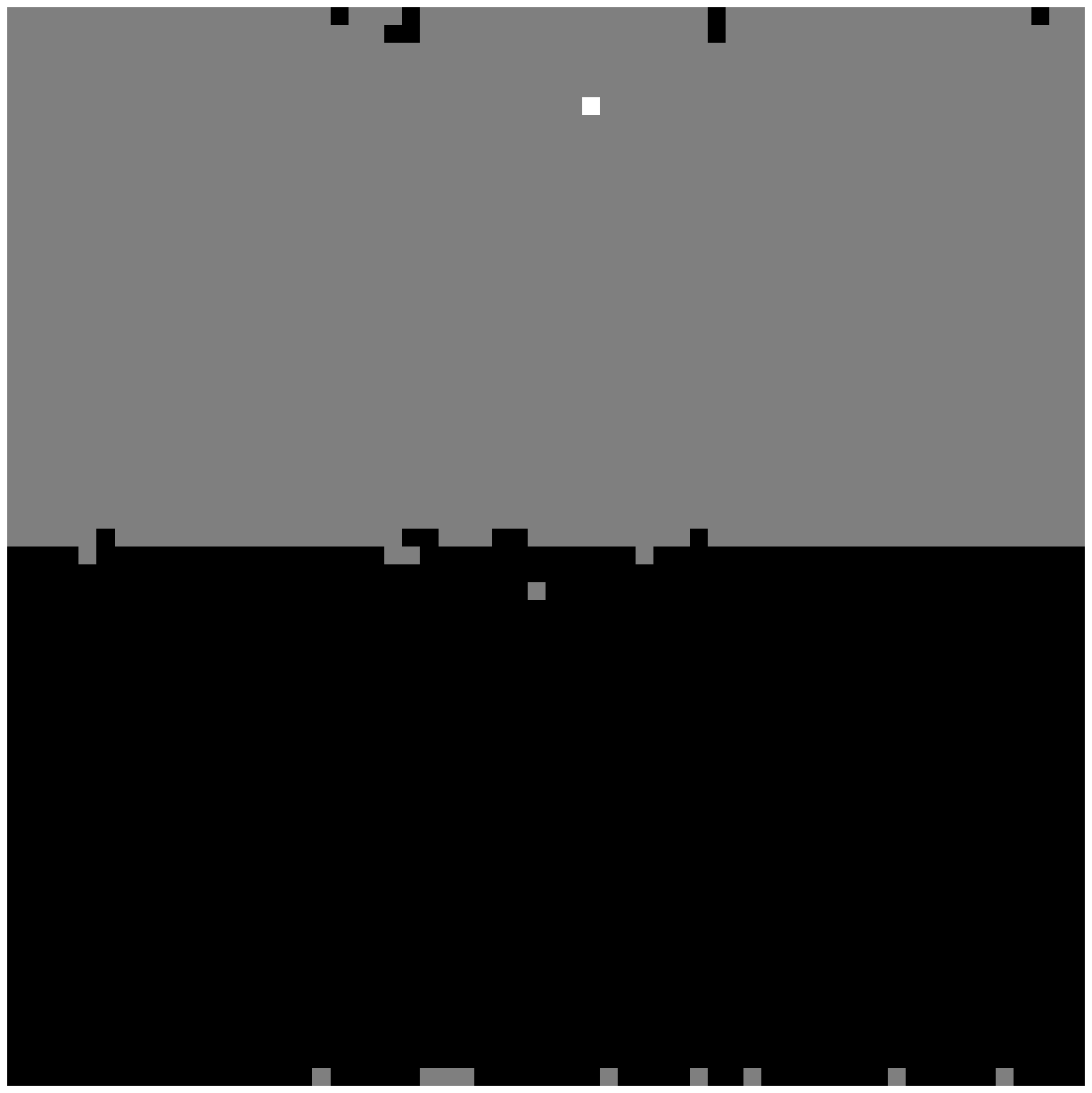} \hfill
    \begin{center} \large{$10^{4}$ MCS} \end{center}
  \end{minipage}
  \hfill \hfill
\begin{minipage}{0.2\textwidth}
  \epsfxsize = \textwidth \epsfysize = 1\textwidth \hfill
  \epsfbox{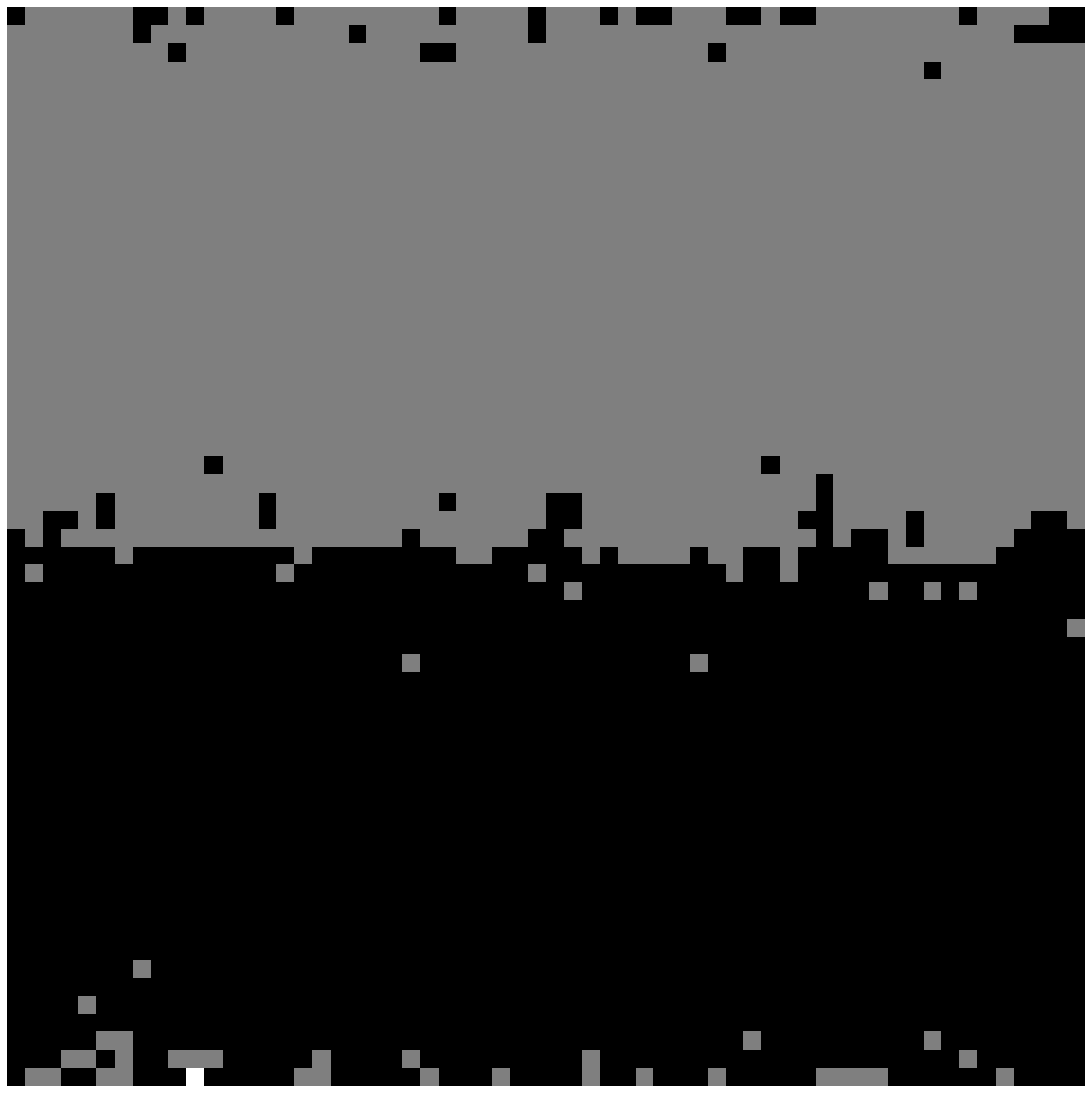} \hfill
    \begin{center} \large{ $10^{5}$ MCS} \end{center}
  \end{minipage}
  \hfill\hfill \vspace{-0.02\textwidth}
\par
\hfill\hfill
\begin{minipage}{0.2\textwidth}
  \epsfxsize = \textwidth \epsfysize = 1\textwidth \hfill
  \epsfbox{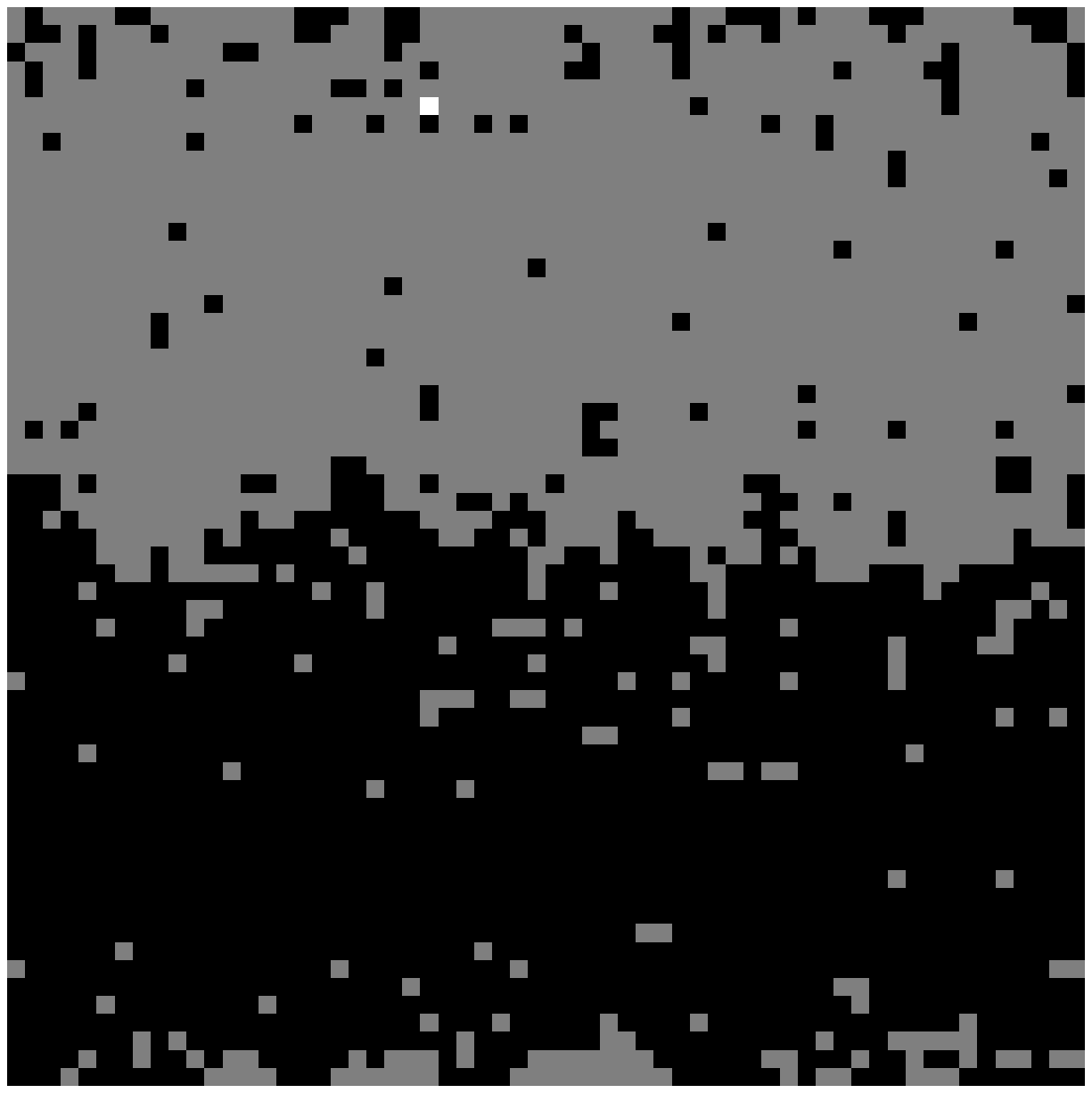}
    \begin{center} \large{$10^{6}$ MCS}  \end{center}
  \end{minipage}
  \hfill \hfill
\begin{minipage}{0.2\textwidth}
  \epsfxsize = \textwidth \epsfysize = 1\textwidth \hfill
  \epsfbox{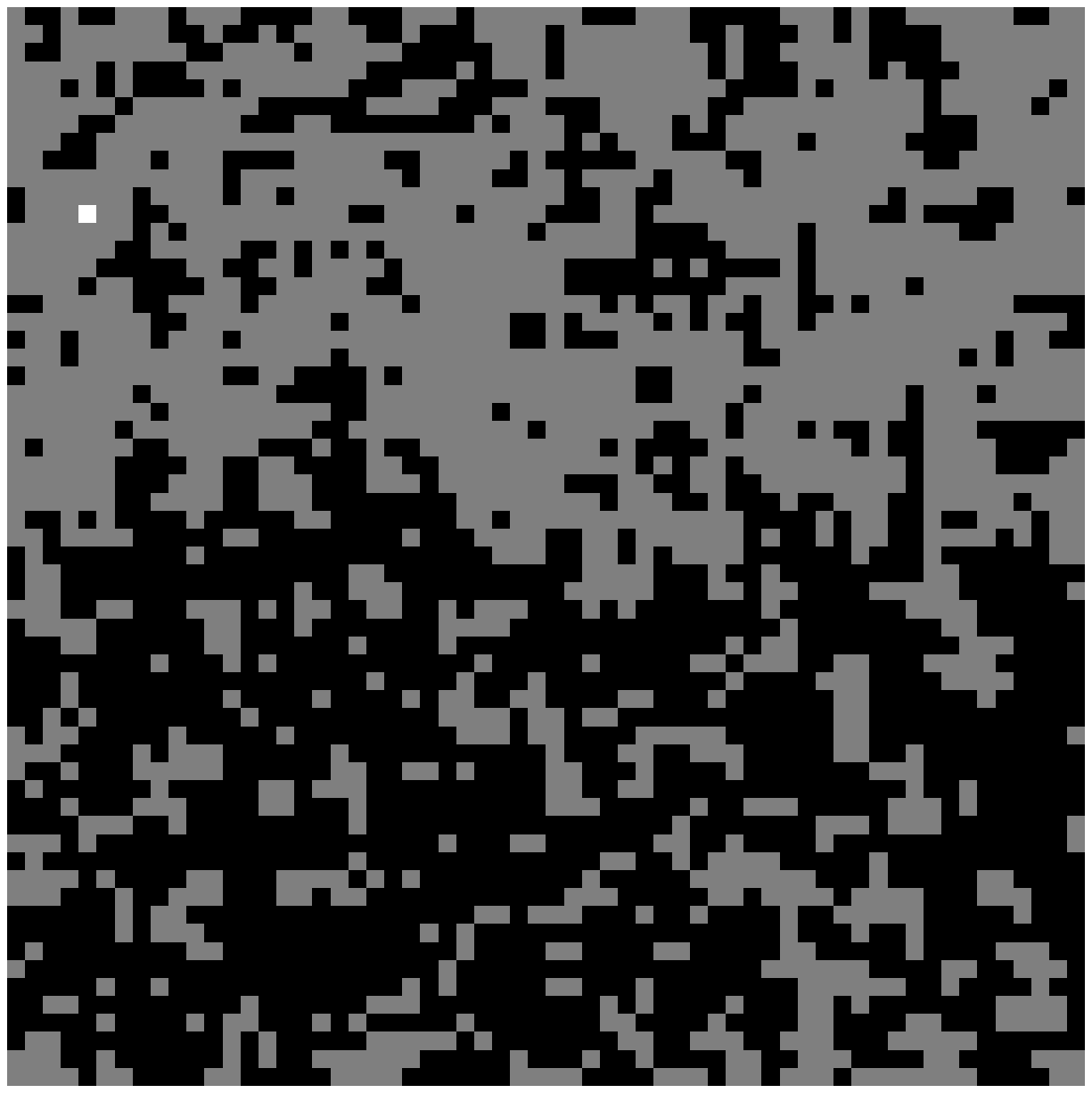}
    \begin{center} \large{$10^{7}$ MCS} \end{center}
  \end{minipage}
  \hfill \hfill
\begin{minipage}{0.2\textwidth}
  \epsfxsize = \textwidth \epsfysize = 1\textwidth \hfill
  \epsfbox{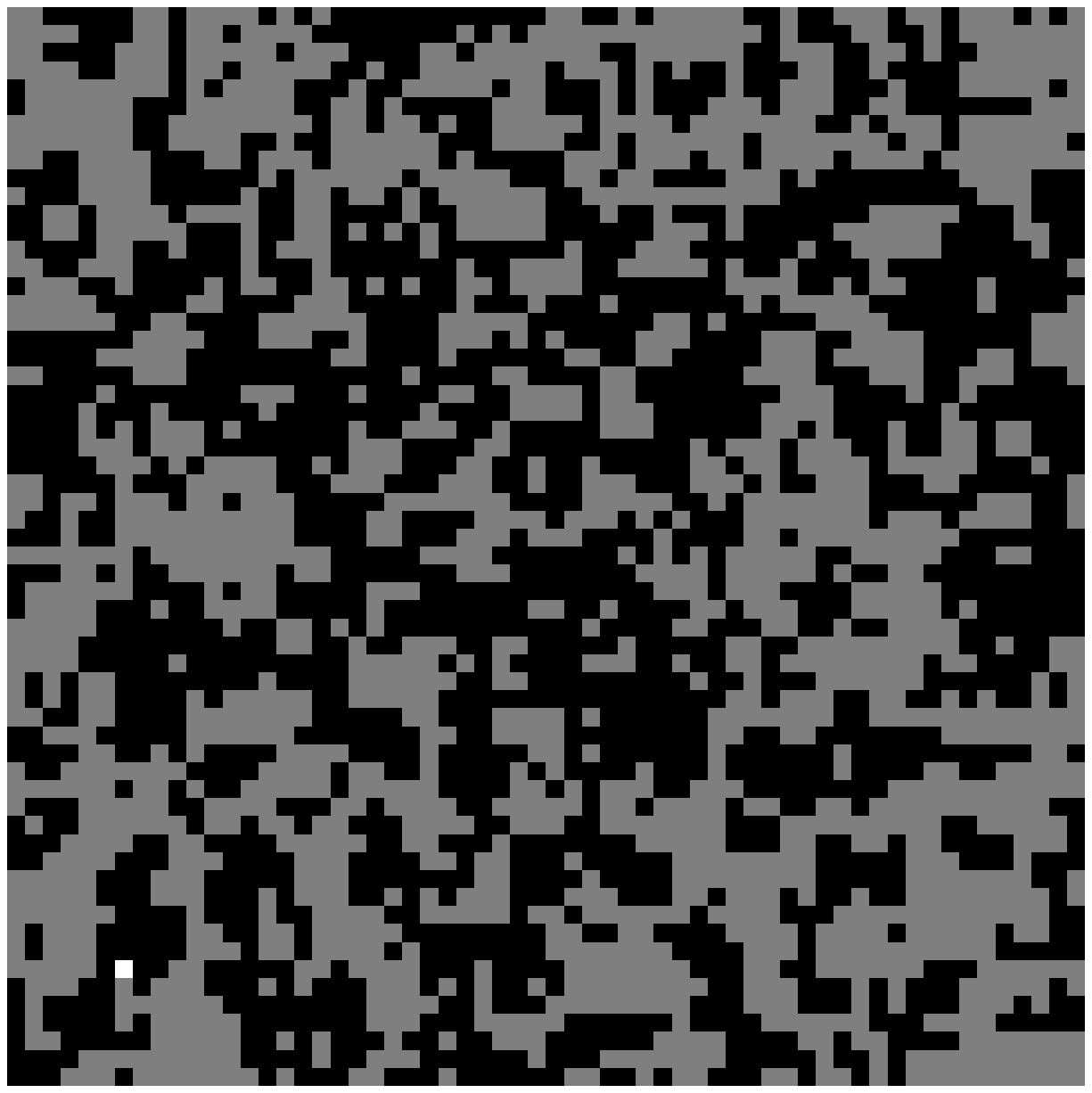}
    \begin{center} \large{ $10^{8}$ MCS} \end{center}
  \end{minipage}
  \hfill \hfill
\vspace{0.03\textwidth}
\caption{Sequence of snapshots showing the disordering process of
a $60 \times 60$
system  with $T= 1.5T_{c}$. The black and gray squares represent the 
two types of particles($\sigma = \pm 1$) and the white square denotes the
vacancy($\sigma = 0$).  The configurations were recorded after  $0,10^4,
10^5,10^6,10^7 \ $and$ \ 10^8$ MCS.  }
\label{p3}
\end{figure}

We conclude that, for temperatures not too close to the critical temperature 
$T_{c}$, the dynamic scaling form of the purely diffusive case, Eqn (\ref
{Asc}), still holds. In two dimensions, we identify the saturation exponent $%
\alpha =2$ and, for a single vacancy, the late crossover exponent $z_{l}=4$.
The scaling function itself, however, develops more curvature in the
intermediate regime, so that it becomes more difficult to extract a pure
power law, $t^{\beta }$. One might be tempted to fit the data to a pure
power law, resulting in a {\em temperature-dependent} exponent $\beta $,
noticeably smaller than $0.5$. However, we believe that there is little
physical support for such a drastic change in scaling properties, especially
in the disordered phase of the Ising model. Instead, we conjecture that $%
\beta =1/2$ is still valid but that $T$-dependent {\em corrections-to-scaling%
} begin to play a stronger role. In support, we checked that the data for $%
T=1.5T_{c}$, in the earlier part of regime II, can be well fitted by a power
law with an exponential correction, i.e., ${\cal A}(L,t)/L^{2}\propto
t^{1/2}\left\{ 1+c_{1}(T)\exp [-c_{2}(T)t/L^{4}]\right\} $. This form is
gleaned from Eqn (\ref{contlim3}), by keeping the leading and first
sub-leading term but allowing for two temperature-dependent fit parameters $%
c_{1}(T)$ and $c_{2}(T)$. Of course, we should anticipate significant
changes in the scaling behavior upon entering the critical region. Since $%
1.1T_{c}$ is ``at the doorstep'' of the latter, some precursors such as an
increased correlation length begin to make their influence felt.

In the following, we comment briefly on some analytic results. First, we can
easily check that the saturation values, ${\cal A}_{sat}(L)$, are consistent
with exact results for the two-dimensional Ising value. Table 1 shows this
comparison for a $60\times 60$ system with a single vacancy, at several
temperatures. ${\cal A}_{sat}^{MC}$ is the {\em measured} saturation value
of the disorder parameter. ${\cal A}_{sat}^{TH}$ is calculated on the basis
of Eqn (\ref{DP}), where $\left\langle {\cal H}\right\rangle $ is the {\em %
exact} bulk value for the average energy of the usual Ising model \cite{MW}.
The effect of the vacancy has been neglected. The agreement is very good,
within the statistical errors of our data. The discrepancies are of course
largest for the last column, with $T$ closest to $T_{c}$.

\begin{figure}[tbp]
\input epsf
\begin{center}
\begin{minipage}{0.5\textwidth}
  \epsfxsize = \textwidth \epsfysize = 1.0\textwidth \hfill
  \epsfbox{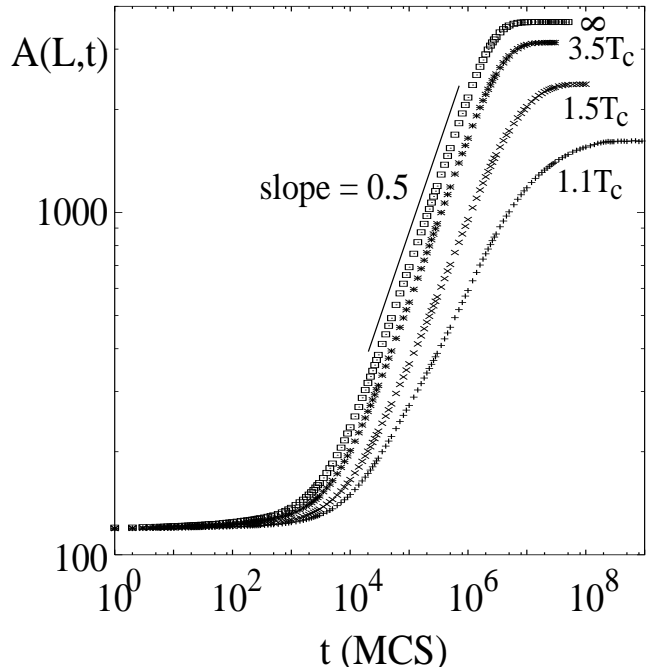}
    \vspace{-.2cm}
\end{minipage}
\end{center}
\caption{$A(L,t)$ vs 
$t$ for $L=60$ with $T= \infty$, $3.5T_{c}$, $1.5T_{c}$ and $1.1T_{c}$. }
\label{p4}
\end{figure}

Finally, since $\left\langle {\cal H}\right\rangle $, and therefore ${\cal A}%
_{sat}(L)$, are extensive, we still have an {\em exact} value for the
saturation exponent $\alpha =d$.

The time-dependence of the disorder parameter poses a much more
difficult problem. It is of course possible to obtain the
temperature-dependent generalization of Eqns (\ref{EoMs}), by averaging the
microscopic master equation, followed by a mean-field approximation and a
naive continuum limit. For a dynamics slightly different from ours, this
procedure was followed in \cite{Puri}. However, these equations are no
longer easily soluble: First, {\em both} of them are inherently nonlinear,
due to interactions. Second, the equation for the local hole density, $\phi (%
{\bf r},t)$, no longer decouples from the magnetization $\psi ({\bf r},t)$,
since the background {\em feeds back} into the motion of the vacancy. Thus,
it is not apparent how a clean time scale separation could emerge from these
equations. As a consequence, we are reduced to numerical solutions or
approximate techniques, such as a high temperature expansion. Preliminary
results, based on the latter, provide some support for our conjecture that $%
\beta =1/2$ is still valid and that larger corrections-to-scaling are
responsible for the curvature in the scaling function \cite{unpub}.

To summarize this section briefly, our data indicate that, for temperatures
not too close to the critical temperature $T_{c}$, the dynamic scaling form
of the purely diffusive case still holds, even though corrections-to-scaling
become more noticeable. In contrast, for $T$ near $T_{c}$ we observe a clear
breakdown of these scaling forms.

\vspace{1.0cm}
\begin{table}
\begin{center}
\caption{Comparison of exact and measured values of the saturated disorder
parameter for several temperatures.}
\vspace{.4cm}
\begin{minipage}{0.4\textwidth}
\begin{tabular}{||c|c|c|c|c||}
$T/T_{c}$ & $\infty $ & $3.5$ & $1.5$ & $1.1$ \\ \hline
${\cal A}_{sat}^{TH}$ & $3600.0$ & $3134.6$ & $2367.0$ & $1602.9\;$ \\ \hline
${\cal A}_{sat}^{MC}$ & $3598.6$ & $3134.0$ & $2368.9$ & $1615.7$ \hfill 
\vspace{0.03pt} \\ 
\end{tabular}
\end{minipage}
\end{center}
\end{table}

\begin{figure}[tbp]
 \input epsf
\begin{minipage}{0.45\textwidth}
  \epsfxsize = \textwidth \epsfysize = .75\textwidth \hfill
  \epsfbox{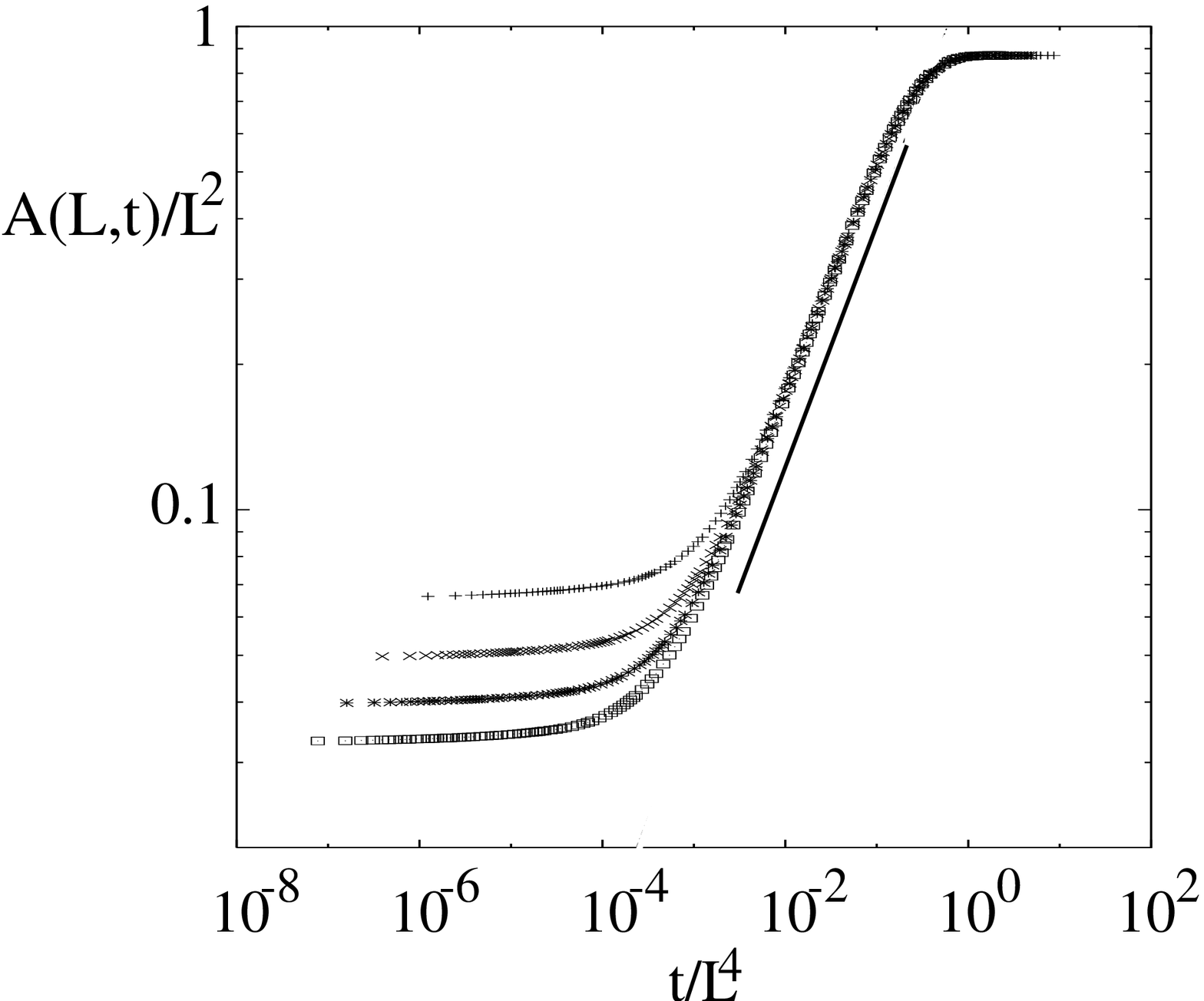}
    \vspace{-.6cm}
    \begin{center} a) \end{center}
\end{minipage}
\hfill
\begin{minipage}{0.45\textwidth}
  \epsfxsize = \textwidth \epsfysize = .75\textwidth \hfill
  \epsfbox{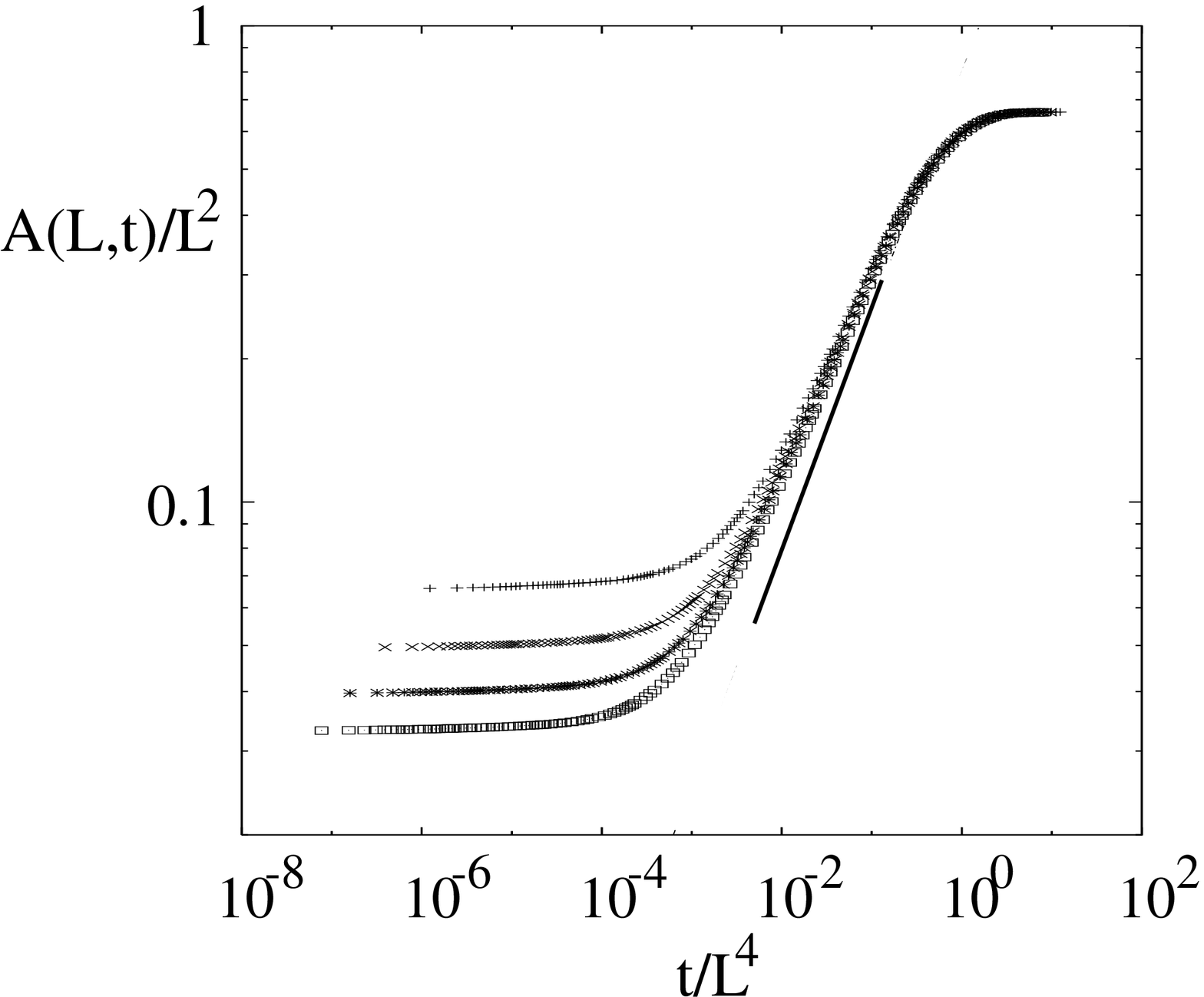}
    \vspace{-.6cm}
    \begin{center} b) \end{center}
\end{minipage}
\vspace{.4cm}
\end{figure}

\begin{figure}[tbp]
\input epsf
\begin{center}
\begin{minipage}{0.45\textwidth}
  \epsfxsize = \textwidth \epsfysize = .75\textwidth \hfill
  \epsfbox{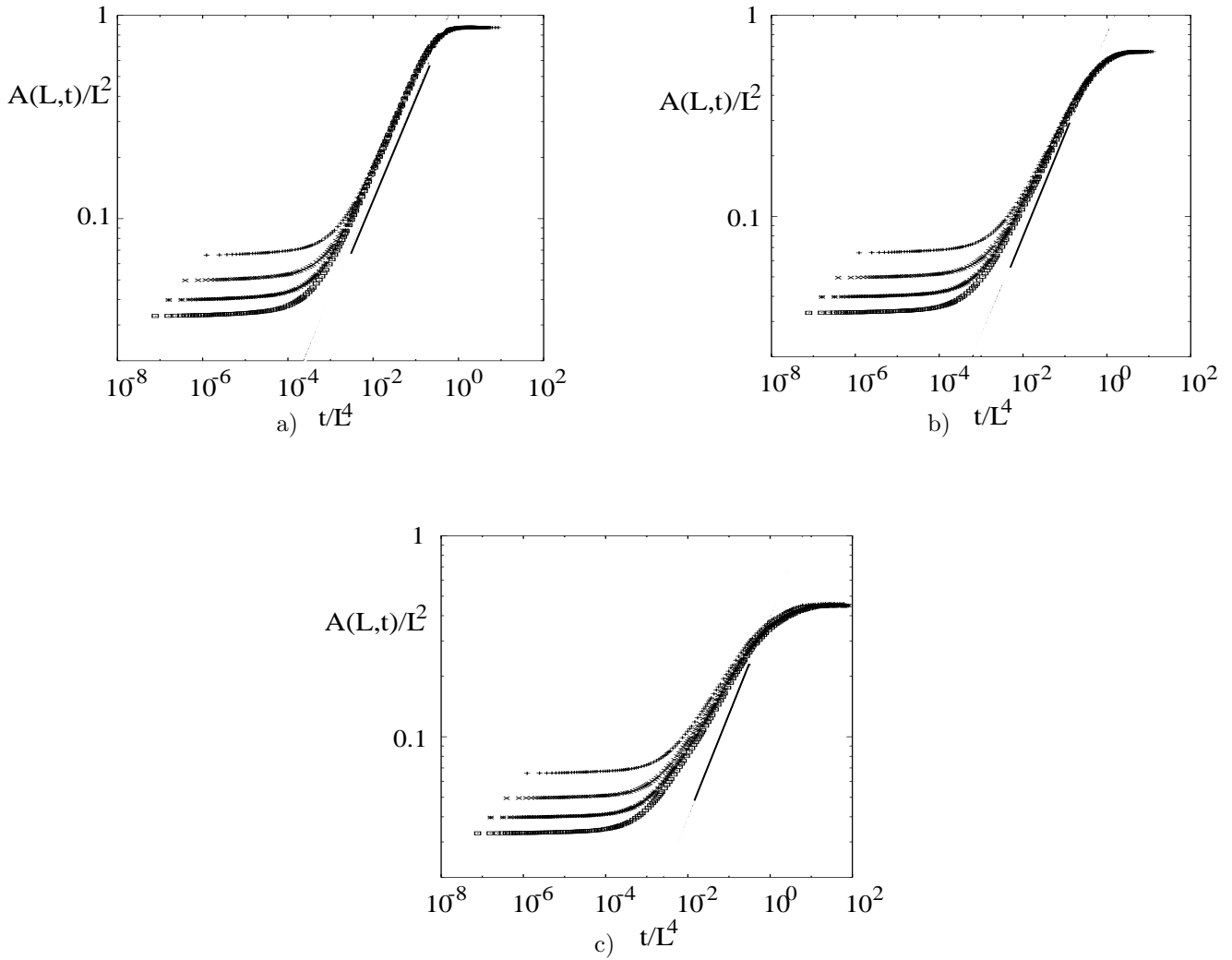}
    \vspace{-.6cm}
    \begin{center} c) \end{center}
\end{minipage}
\end{center}
\caption{The scaling plot of $A(L,t)/L^{2}$ vs 
\ $t/L^{4}$ for $L=30-60$ with (a) \ $T= 3.5T_{c}$, (b) \ $T= 1.5T_{c}$ and 
(c) \ $T= 1.1T_{c}$. The straight reference lines have slope $0.5$.}
\label{p4}
\end{figure}

\section{Concluding Remarks}

To summarize, we have analyzed the vacancy-driven disordering process of an
initially phase segregated binary system, which is rapidly heated to a
temperature $T$ above criticality. The number of vacancies scales with
system size as $L^{\gamma }$, where the vacancy exponent $\gamma $ is a
parameter. To quantify the evolution of the system, we measure the number of
broken bonds, ${\cal A}(L,t)$, as a function of time. This ``disorder
parameter'' allows us to identify three temporal regimes, distinguished by
the distribution of the vacancies through the system. Our key result is that
the late stages of this process exhibit dynamic scaling. A set of exponents $%
\left\{ z_{e},z_{l},\alpha ,\sigma ;\beta \right\} $ can be defined,
characterizing, respectively, the system size dependence of two crossover
times, the final saturation value of the disorder parameter, and its
amplitude in the intermediate regime. The temporal growth of ${\cal A}$
during the latter regime is captured by the exponent $\beta $. If the final
temperature is infinite, i.e., $T=\infty $, the motion of the vacancies is a
simple Brownian random walk, and all indices can be computed analytically,
in excellent agreement with the data. In particular, we find that the
typical time scale $t_{l}\sim L^{z_{l}}$, which controls the crossover
between increasing disorder and saturation, is set by $z_{l}=2+d-\gamma $,
and thus depends explicitly on both the space and vacancy dimensionalities, $%
d$ and $\gamma $. Measurements of $z_{l}$ can therefore provide information
about the vacancy distribution in a sample. In the most familiar case,
standard vacancy diffusion in solids, the number of vacancies is extensive ($%
\gamma =d$), so that the well-known result $z_{l}=2$ is reproduced \cite
{basic MS}. In contrast, in two dimensions we find $z_{l}=4$ for a single
vacancy ($\gamma =0$) and $z_{l}=3$ if the defects are generated at the
initial interface. For {\em finite} temperature $T\gtrsim 1.5T_{c}$, we
observe that these scaling forms still hold. Even though interparticle
interactions now play a role, correlations in the system are still
short-ranged, so that the vacancy still performs a random walk if viewed on
a length scale which exceeds the correlation length, $\xi (T)$. As $T$
approaches $T_{c}$, however, $\xi (T)$ reaches $O(L)$ so that the simple
random walk scenario must break down. In our simulations, significant
deviations from the high-temperature scaling form appear already for $%
T=1.1T_{c}$, where $\xi (T)\simeq 6$, in units of the lattice spacing \cite
{MW}.

All the investigations above are focused on upquenches to the disordered
state. A natural extension of this project is to set the final temperature
below criticality. Needless to say, new questions and interesting phenomena
can be expected. In particular, the final equilibrium state is still phase
separated, so that all the interfacial properties come into play. With a
single vacancy in a finite system, there may be a further separation of
times scales. We expect that, sooner after the early stage, the
``intrinsic'' density profile of the interface can be built. At this point,
the interfacial width, $w$, is most likely controlled by $\xi $. On the
other hand, at sufficiently long times, the capillary waves will surely make
their presence felt. In the $d=2$ case, the interface is always rough, so
that we can always expect the latter crossover to occur. How, and if, $w(\xi
,L;t)$ scales will be of great interest. Besides dynamics and timescales, we
could study the equilibrium probability profile of the {\em vacancy}, i.e.,
where the vacancy spends most of its time. In case more than one vacancy is
present, this profile should map into a density profile for the vacancies.
In either case, we expect the vacancies to be trapped, to a greater or
lesser extent, at the interface. Now, if more and more vacancies are added
to system, the interface profile should be altered, since a preponderance of
vacancies may significantly modify the interfacial energy. For $d=3$ Ising
models, there are further interesting phenomena, associated with roughening
transitions. Given that interfacial energies should depend on vacancy
concentrations, how the locations (if not the nature) of such transitions
are affected is a natural question. Finally, it would be extremely
interesting to study the dynamic content of these systems, such as scaling
properties when the upquench is set at a roughening temperature.

We conclude with a few remarks on related problems. It is obviously
interesting to test the range of universality of the high-temperature
scaling forms, with respect to changes in the boundary conditions and the
microscopic dynamics. Returning to $T=\infty $, we have investigated several
cases. First, replacing periodic by reflecting boundary conditions changes
only trivial amplitudes. A more significant modification is the addition of
a bias $E$, such as a gravitational or electric field \cite{ats}: if all
particles carry equal charge while the vacancy is neutral, vacancy moves
along a selected direction will be suppressed. In this case, the boundary
conditions determine the nature of the resulting steady state. For
reflecting boundary conditions, it is described by a simple {\em Boltzmann
factor}, incorporating a gravitational or an electrostatic potential. Here,
a new scaling variable, $\Delta \equiv LE$, emerges which characterizes the
potential difference between opposing boundaries. In contrast, in the
periodic case the steady state is a {\em non-equilibrium} one. Moreover, it
is translation-invariant, so that all dependence on $E$ disappears in a
suitably chosen {\em co-moving} frame! Remarkably, however, these
boundary-induced differences appear {\em only} in the scaling functions,
leaving us with a {\em universal} set of scaling exponents \{$%
z_{e},z_{l},\alpha ,\sigma ;\beta $\} \cite{ats}.

Finally, we should add a few remarks on the early regime. As pointed out
above, for an infinite system with a single vacancy, this is the {\em only }%
regime. First, notice that only the phenomenon of interfacial destruction is
relevant, since bulk disordering would take infinitely long times. Next,
considering the effective range of a Brownian walker, it is clear that the
size of the disordered region grows at most with $t^{1/2}$. In more detail,
we see that there are two distinct types of disorder. Loosely, we will label
them ``transverse'' and ``longitudinal.'' The first type is associated with
a monolayer-like disorder from the initially sharp interface, spreading from
where the vacancy began its journey. In particular, each time the vacancy
crosses the plane at a new location, a particle is moved into the ``wrong''
phase. Thus, we may expect transverse disorder to spread like $t^{1/2}$. The
second type is a measure of the width of the interfacial profile. For this
to occur, the vacancy must ``carry'' e.g., a black particle deeper and
deeper into the sea of white ones. Thus, the return probability of the
walker plays a crucial role. Only in $d=2$ is this probability both relevant
and non-trivial. Applying the well-known results of random walks \cite
{Hughes}, we find that ``longitudinal'' disorder grows as $\left( \ln
t\right) ^{1/2}$\cite{TZ}.

Even though our model is very simple, it forms the basis for the description
of a large variety of related systems. Moreover, it is truly gratifying that
considerable analytic progress is possible for a problem that is both {\em %
nonlinear} and {\em time-dependent}. Work is in progress to probe the effect
of fluctuations on the structural decomposition process at a more
microscopic level \cite{astz}.

\begin{center}
{\bf {ACKNOWLEDGMENTS} }
\end{center}

The authors gratefully acknowledge fruitful collaborations with T.
Aspelmeier, G. Korniss, and Z. Toroczkai. We also thank J. Beauvais, M.
Grant, T. Newman, S. Redner and B. Taggart for helpful discussions. This
work was supported in part by the National Science Foundation through the
Division of Materials Research. \newline

\newpage
\begin{center}
{\bf {APPENDIX: POISSON RESUMMATION} }
\end{center}

In this Appendix, we add a few details of the Poisson resummation procedure.
Or starting point is Eqn (\ref{Acl1}). Dropping the constant prefactor, we
focus on the expression in the brackets, with $\zeta \equiv \epsilon t$: 
\begin{equation}
F(\zeta )\equiv 1-\frac{8}{\pi ^{2}}\sum_{n=1}^{\infty }e^{-\zeta
(2n-1)^{2}}/(2n-1)^{2}\;  \label{F1}
\end{equation}
This sum converges rapidly in the limit $\zeta \rightarrow \infty $. We seek
a form which converges equally well in the limit $\zeta \rightarrow 0$. We
first recast this sum in terms of a much simpler one, namely, 
\begin{equation}
S(\lambda )\equiv \sum_{n=1}^{\infty }e^{-(\lambda n)^{2}}.  \label{S}
\end{equation}
We begin by noting that $\sum_{1}^{\infty }1/(2n-1)^{2}=\pi ^{2}/8$ so that 
\begin{eqnarray}
F(\zeta ) &=&\frac{8}{\pi ^{2}}\sum_{n=1}^{\infty }\frac{1-e^{-\zeta
(2n-1)^{2}}}{(2n-1)^{2}}  \nonumber \\
&=&\frac{8}{\pi ^{2}}\zeta \int_{0}^{1}ds\sum_{n=1}^{\infty }e^{-\zeta
(2n-1)^{2}s}
\end{eqnarray}
It is easy to recognize the sum in this expression as the ``odd'' terms in $%
S(\lambda )$, with $\lambda =\sqrt{s\zeta }$. This suggests that we should
recast $S(\lambda )$ in the form 
\begin{eqnarray*}
S(\lambda ) &=&\sum_{n=1}^{\infty }e^{-(2\lambda n)^{2}}+\sum_{n=1}^{\infty
}e^{-(\lambda (2n-1))^{2}} \\
&\equiv &S_{e}(\lambda )+S_{o}(\lambda )
\end{eqnarray*}
where the ``even'' terms just reproduce $S(\lambda )$, via $S_{e}(\lambda
)=S(2\lambda )$. Summarizing so far, we obtain 
\begin{eqnarray}
F(\zeta ) &\equiv &\frac{8}{\pi ^{2}}\zeta \int_{0}^{1}ds\ S_{o}(\sqrt{\zeta
s})  \nonumber \\
&=&\frac{8}{\pi ^{2}}\zeta \int_{0}^{1}ds\ \left[ S(\sqrt{\zeta s})-S(2\sqrt{%
\zeta s})\right] \;  \label{F2}
\end{eqnarray}
Thus, it is sufficient for us to resum $S(\lambda )$. We recall the Poisson
resummation formula \cite{Jones}, for a function $f\left( x\right) $,
defined on $0\leq x<\infty $ with $\lim_{x\rightarrow \infty }f(x)=0$ and $%
\int_{0}^{\infty }dx\ f(x)$ finite: 
\begin{equation}
\frac{1}{2}f(0)+\sum_{n=1}^{\infty }f(n\lambda )=\lambda ^{-1}\left[ \frac{1%
}{2}\tilde{f}(0)+\sum_{m=1}^{\infty }\tilde{f}(\frac{2\pi m}{\lambda }%
)\right]  \label{Pf}
\end{equation}
where $\tilde{f}$ is the Fourier cosine transform of $f$: 
\[
\tilde{f}(\alpha )\equiv 2\int_{0}^{\infty }dx\ f(x)\cos (\alpha x)\ . 
\]
From Eqns (\ref{S}), we can immediately read off $f(x)=\exp (-x^{2})$ for
our case whence 
\[
\tilde{f}(\alpha )=\sqrt{\pi }\exp (-\alpha ^{2}/4)\ . 
\]
Thus, we obtain 
\begin{equation}
S(\lambda )=-\frac{1}{2}+\lambda ^{-1}\left\{ \frac{\sqrt{\pi }}{2}+\sqrt{%
\pi }\sum_{m=1}^{\infty }e^{-(m\pi /\lambda )^{2}}\right\}  \label{PS}
\end{equation}
Clearly, the original form of $S(\lambda )$, Eqn (\ref{S}) converges rapidly
for large $\lambda $, while the alternate form presented here converges
rapidly for $\lambda \rightarrow 0$. Inserting our result into Eqn (\ref{F2}%
) and performing the integration results in 
\begin{equation}
F(\zeta )=\frac{4}{\pi ^{3/2}}\sqrt{\zeta }\left\{ 1+2\sum_{m=1}^{\infty
}(-1)^{m}\left[ e^{-u_{m}^{2}}-u_{m}\Gamma (\frac{1}{2},u_{m})\right]
\right\} \;  \label{F3}
\end{equation}
where $u_{m}\equiv \pi m/\sqrt{8\epsilon t}$. This leads immediately to our
result for the disorder parameter, Eqn (\ref{contlim3}).

The Poisson resummation of the magnetization profile, Eqn (\ref{psisol}), is
a little more involved, but follows an analogous series of steps. In this
case, the problem can be reduced to resumming 
\[
\Sigma (\tilde{y},\zeta )\equiv \sum_{n=1}^{\infty }\frac{\sin (n\tilde{y})}{%
n}e^{-\zeta n^{2}}, 
\]
where $\tilde{y}\equiv 2\pi y/L$, since we can rewrite 
\[
\psi ({\bf r},t)=\frac{4}{\pi }\left[ \Sigma (\tilde{y},\zeta )-\frac{1}{2}%
\Sigma (2\tilde{y},4\zeta )\right] \text{ .} 
\]
The resummation leads to 
\[
\Sigma (\tilde{y},\zeta )=-\frac{\tilde{y}}{2}+\frac{\pi }{2}%
\mathop{\rm erf}%
\left( \frac{\tilde{y}}{2\sqrt{\zeta }}\right) +\frac{\pi }{2}%
\sum_{m=1}^{\infty }\left[ 
\mathop{\rm erf}%
\left( \frac{\tilde{y}+2\pi m}{2\sqrt{\zeta }}\right) +%
\mathop{\rm erf}%
\left( \frac{\tilde{y}-2\pi m}{2\sqrt{\zeta }}\right) \right] 
\]
so that we finally arrive at Eqn (\ref{psisolP}).


\end{document}